\documentclass[aps,notitlepage,nofootinbib,longbibliography,twocolumn,superscriptaddress,10pt]{revtex4-2}

\usepackage[utf8]{inputenc}
\usepackage[T1]{fontenc} 
\usepackage{bm}
\usepackage{bbm}
\usepackage{ulem}
\usepackage[dvipsnames]{xcolor} 
\usepackage{wrapfig}
\usepackage{amsmath}
\usepackage{amssymb}
\usepackage{mathtools}
\usepackage{hyperref}
\usepackage{tikz}
\newcommand{\be}{\begin{equation}}
\newcommand{\ee}{\end{equation}}
\renewcommand{\S}{\mathcal{S}}
\newcommand{\bx}{\bm{x}}
\newcommand{\Z}{\mathbb{Z}}
\newcommand{\vphi}{\varphi}
\newcommand{\C}{\mathcal{C}}
\newcommand{\B}{\mathcal{B}}
\newcommand{\Q}{\mathcal{Q}}
\renewcommand{\L}{\mathcal{L}}
\newcommand{\p}{\partial}

\begin{document}
\title{Gapped boundaries of (3+1)d topological orders}
\author{Zhu-Xi Luo}
\affiliation{Department of Physics, Harvard University, Cambridge, MA 02138, USA}

\begin{abstract}
Given a gapped boundary of a (3+1)d topological order (TO), one can stack on it a decoupled (2+1)d TO to get another boundary theory. Should one view these two boundaries as ``different''? A natural choice would be no. Different classes of gapped boundaries of (3+1)d TO should be defined modulo the decoupled (2+1)d TOs. But is this enough? 

We examine the possibility of {\it coupling} the boundary of a (3+1)d TO to additional (2+1)d TOs or fractonic systems, which leads to even more possibilities for gapped boundaries. Typically, the bulk point-like excitations, when touching the boundary, become excitations in the added (2+1)d phase, while the string-like excitations in the bulk may end on the boundary but with endpoints dressed by some  other excitations in the (2+1)d phase. For a good definition of ``class'' for gapped boundaries of (3+1)d TO, we choose to quotient out the different dressings as well.

We characterize a {\it class} of gapped boundaries by the string-like excitations that can end on the boundary, whatever their endpoints are. A concrete example is the (3+1)d bosonic toric code. Using group cohomology and category theory, three gapped boundaries have been found previously: rough boundary, smooth boundary and twisted smooth boundary. We can construct many more gapped boundaries beyond these, which all naturally fall into two classes corresponding to whether the $m$-string can or cannot end on the boundary. According to this classification, the previously found three boundaries are grouped as \{rough\}, \{smooth, twisted smooth\}. 
For a (3+1)d TO characterized by a finite group $G$, different classes correspond to different normal subgroups of $G$. 
We illustrate the physical picture from various perspectives including coupled layer construction, Walker-Wang model and field theory.
\end{abstract}

\maketitle
\tableofcontents

\section{Introduction}

To this date, the gapped boundaries and interfaces of (2+1)d topological phases of matter have been well-studied using the formalisms of Lagrangian subgroups, anyon condensations, Frobenius algebras and tunneling matrices, etc. \cite{Juven,Levin,ChaoMing_bdry,PhysRevLett.89.181601,bais2003hopf, BAIS2007552,PhysRevB.79.045316, kitaev2012models,KONG2014436,Hung_Wan,hung2015generalized, Bernevig1, Bernevig2,  hu2018boundary,PhysRevB.96.165138,PhysRevLett.114.076402}, see \cite{Fiona} for a review.  Although important progresses such as \cite{PhysRevB.105.155137,Hu_2022} are still being made, the community has gradually shifted its attention towards (3+1)-dimensions. 

There are several works on the boundaries theories of (3+1)d topological orders, which we will review in subsection \ref{subsec:previous}. The subject is, however, not closed. One of the most important open questions is a sharper definition of different classes of gapped boundaries. Given a gapped boundary of a (3+1)d topological order, if one stacks another (2+1)d topological order on top that does not interact with the (3+1)d theory or its boundary, the new boundary theory will include additional deconfined excitations from the added (2+1)d theory. Naïvely, before and after the stacking, the gapped boundaries have different contents of topological excitations and should therefore be viewed as different. This difference is, however, artificial and not intrinsic to the (3+1)d topological order itself. It is therefore natural to define a class of gapped boundaries of (3+1)d topological orders by modding out these redundant effects of the added decoupled (2+1)d topological orders. This understanding, although have not yet been explicitly spelled out (to our knowledge) in literature, is a folklore among experts. 

In this work, we examine the question of whether quotienting out decoupled (2+1)d theories is enough. We find a large family of unexplored gapped boundaries  with nontrivial surface excitations that are not present in the bulk. Such gapped boundary theories are constructed from \textit{coupling} an additional (2+1)d theory to the (3+1)d bulk. The (2+1)d theory can be a topological order, or something more exotic such as a fractonic theory. The resultant gapped boundaries are typically anomalous, in the sense that the surface excitations and anti-excitations may not be freely separable from each other. In the simplest example of (3+1)d bosonic $\Z_2$ toric code, we find numerous new gapped boundaries, which neatly fall into two different classes represented by the two subgroups of $\Z_2$. The two classes correspond to whether the $m$-string can or cannot end on the boundary, regardless of what lives at its endpoint. In general, we distinguish various classes of gapped boundaries of (3+1)d topological orders based on the \textit{different types strings in the bulk that are allowed to end on the boundary}. 
For convenience of the readers, we will present a preview of the main ideas in subsection \ref{subsec:idea}. 

Before moving on to more details, we will comment on another motivation of this work.  In (3+1)-dimensions, more exotic physics can be present on top of topological orders, such as restricted, fractionalized mobilities of excitations in fracton phases of matter \cite{Chamon,BRAVYI2011839, Haah,Yoshida,PhysRevB.92.235136,Vijay_original, Rahul_review,Pretko_review} which are generally described by tensor gauge theories \cite{Cenke,gu2006lattice,PhysRevD.81.104033,GU201290,rasmussen2016stable,Pretko1, Pretko2,KevinKim,bulmash2018generalized,PhysRevX.9.031035, PhysRevB.101.085106,Yau,Seiberg,YOU2020168140,SS1,SS2,SS3,SS4,PhysRevResearch.2.023249,fontana2021field}. Recently, references \cite{Daniel, Me, Wenjie} discussed the gapped boundary theories for certain fracton phases of matter. However, in order to have a full understanding of all possible gapped boundaries for fracton phases of matter, a better understanding of the topological phases of matter, which have simpler algebraic structures, is required. 

The remaining of the paper is organized as follows. We will review the existing works on the boundary theory of topological phases in (3+1)d in subsection \ref{subsec:previous}, and present a preview of the main construction of general gapped boundaries in subsection \ref{subsec:idea}, using the example of (3+1)d $\Z_2$ toric code. Then in sections \ref{sec:layer}, \ref{sec:WW} and \ref{sec:field}, we will use different formalisms to illustrate the idea in more details, including the coupled layer construction, the Walker-Wang-type model, and quantum field theory. We will also discuss the anomaly inflow  \cite{CALLAN1985427} in the field theory language to reconstruct the (3+1)d bulk theories from the boundary theories. Section \ref{sec:EBL} demonstrates an example where the (2+1)d $\Z_N$ plaquette model, which has subsystem symmetries, is used to construct a gapped boundary for the (3+1)d toric code. Finally in section \ref{sec:discussion} we discuss the expectations for general (3+1)d topological orders beyond toric code, and comment on other issues.

\subsection{Previous works}
\label{subsec:previous}

Prior to this work, three different types of gapped boundaries for the (3+1)d bosonic $\Z_2$ toric code have been discussed in the literature, called the rough boundary, smooth boundary, and twisted smooth boundary. (The names rough and smooth originated from the shape of the boundary Hamiltonians written in terms of Pauli matrices.)

Ref. \cite{Yidun} studied the gapped boundaries of twisted gauge theories by systematically constructing the boundary Hamiltonians on the lattice. If the bulk theory is characterized by a gauge group $G$ and a 4-cocycle $\alpha \in H^4[G,U(1)]$ in the fourth cohomology group of $G$ over $U(1)$, a gapped boundary is determined by a subgroup $K\subset G$ and a 3-cochain $\beta \in C^3[K,U(1)]$ in the third cochain group of $K$ over $U(1)$.  In the case of (3+1)d $\Z_2$ bosonic toric code, $G=\Z_2$ and $\alpha=1$ is trivial, there are two subgroups $K_1=\Z_1$ and $K_2=\Z_2$. $C^3[K_1,U(1)]$ is trivial, giving rise to the rough boundary condition, while the two elements in $C^3[K_2,U(1)]$ lead to the smooth and twisted smooth boundary conditions respectively. 

More recently, ref. \cite{Janet} studied the gapped boundaries of $\Z_2$ toric code from the perspective of string condensations and Lagrangian algebras in the modular 2-category. In addition, they provided a coupled layer construction picture for the three different gapped boundaries, which we will also use and generalize in section \ref{sec:layer}. 

Another way of obtaining gapped boundary conditions was described in ref. \cite{PhysRevX.8.031048}, where instead of breaking gauge symmetries at the boundary, one enhances the gauge symmetries. This in principle can lead to more than three gapped boundaries for the $\Z_2$ toric code. The new gapped boundaries constructed in our work include examples that cannot be obtained through the symmetry enhancing procedure described in ref. \cite{PhysRevX.8.031048} (see for example sections \ref{subsec:Ising} and \ref{sec:EBL}), however, at this stage it is unclear whether all the gapped boundaries available through symmetry enhancing can be obtained from our setup and future investigations are needed to pin down the relationships.

Ref. \cite{Shinsei} discussed the gapless boundary theories for the 3d $\Z_N$ toric code as well as the case with a $\Theta$ term. The authors also studied the bulk-boundary correspondence by matching the modular $S$ and $T$ matrices computed from the boundary field theories with those computed in the bulk. Ref. \cite{Kong_2017} argued using category theory that given an $n$-dimensional gapped boundary theory, one can uniquely reconstruct its $(n+1)$-dimensional anomaly-free topological order by taking the center of the boundary theory.  We will discuss the bulk-boundary correspondence from another perspective using anomaly inflow \cite{CALLAN1985427}, see sections \ref{subsec:inflow} and \ref{subsec:EBL_field}. Typically the anomalies of the boundary theories will be canceled by (3+1)d symmetry protected topological phases.  

Below for convenience we will use (2+1)d and 2d interchangeably, and similar for (3+1)d and 3d.

\subsection{Main idea}
\label{subsec:idea}

\begin{figure}[htbp]
\centering
\begin{tikzpicture}[scale=0.7]
\draw[ForestGreen,thick] (2.5,0.5)--(2.5,3.15);
\draw (0,0)--(0,3)--(3,3)--(4,4)--(1,4)--(0,3);
\draw (3,3)--(3,0);
\draw (4,4)--(4,1);
\draw[dashed] (1,4)--(1,1);
\draw (0,4.3)--(3,4.3)--(4,5.3)--(1,5.3)--(0,4.3);
\node at (5.1 ,2.2) {3d TC};
\node at (5.1 ,4.7) {2d TO};
\node at (1.3,3.5) {\textcolor{blue}{$e$}};
\node at (1.3,4.75) {\textcolor{blue}{$b$}};
\node at (2.5,3.5) {\textcolor{ForestGreen}{\small $m$}};
\node at (2.5,4.75) {\textcolor{ForestGreen}{\small $f$}};
\draw[blue] (1.3,3.5+0.625) ellipse (0.2 and 1);
\draw[ForestGreen] (2.5,3.5+0.625) ellipse (0.3 and 1);
\end{tikzpicture}
\begin{tikzpicture}[scale=0.7]
\draw[ForestGreen,thick] (1.3,3.15) .. controls (1.7,1.8) and (2.1,1.8)  .. (2.5,3.15);
\draw (0,0)--(0,3)--(3,3)--(4,4)--(1,4)--(0,3);
\draw (3,3)--(3,0);
\draw (4,4)--(4,1);
\draw[dashed] (1,4)--(1,1);
\draw (0,4.3)--(3,4.3)--(4,5.3)--(1,5.3)--(0,4.3);
\node at (1.3,3.5) {\textcolor{ForestGreen}{\small $m$}};
\node at (1.3,4.75) {\textcolor{ForestGreen}{\small $f$}};
\node at (2.5,3.5) {\textcolor{ForestGreen}{\small $\bar{m}$}};
\node at (2.5,4.75) {\textcolor{ForestGreen}{\small $\bar{f}$}};
\draw[ForestGreen] (1.3,3.5+0.625) ellipse (0.3 and 1);
\draw[ForestGreen] (2.5,3.5+0.625) ellipse (0.3 and 1);
\end{tikzpicture}
\caption{General procedure of finding a gapped boundary for the 3d $\Z_2$ toric code. Left: condense the composite of the charge $e$ with a bosonic excitation $b$ in a 2d gapped phase of matter. The $m$-string can end on the surface with the endpoint being an excitation $f$, if $f$ has mutual semionic statistics with $b$. There can be no $f$ or multiple $f_i$'s. Right: The $f$ excitations, which can be deconfined in the original decoupled 2d TO, become confined when the composite $b\otimes e$ is condensed, because they are connected by a $m$-string in the bulk.}
\label{fig:idea}
\end{figure}
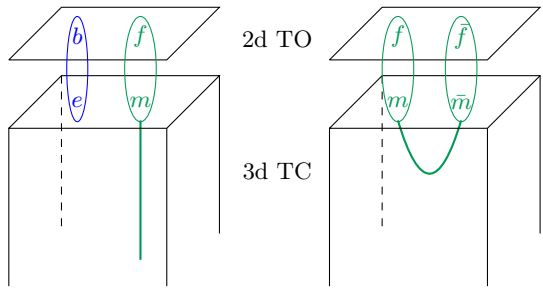

Even in the simplest case of $\Z_2$ toric code, we find significantly many more options for gapped boundaries. We summarize the construction in figure \ref{fig:idea} using this example. 

For any 2d gapped phase of matter with a bosonic excitation $b$ (which can be trivial), one can condense the composite of $b$ with the charge excitation $e$ in the toric code. This is allowed because the composite is again bosonic. If there exists another excitation $f$ in the 2d theory that has mutual semionic statistics with $b$, then the $m$-string in the bulk is allowed to end at the boundary, with its endpoint being $f$.  The pair $m\otimes f$ commutes with $e\otimes b$ and is therefore not confined by the condensate. However, if one creates a pair of $(f\otimes m, \bar{f}\otimes \bar{m})$, the excitations and anti-excitations cannot be freely separated from each other since the $m$ and $\bar{m}$ are connected by a energetically costly $m$-string in the bulk. In this sense, the boundary is anomalous.  We have deliberately kept the 2d phase to be general in the descriptions above: while most of the examples in our paper require the 2d theory to be a topological phase of matter, one can also easily construct examples where the 2d theory is not topological, see for example section \ref{sec:EBL}, where the $\Z_N$ plaquette model is used and the surface excitations need to satisfy additional mobility constraints. 

In general there can be more than be multiple $f_i$'s which can all serve as the endpoints of the $m$-string on the boundary. They may have nontrivial mutual statistics with each other. It can also happen that there does not exist any $f$ in the 2d phase that satisfies the requirement. In this case, the $m$-string in the bulk is not allowed to end at the boundary. All gapped boundaries of toric code fall into two \textit{classes} corresponding to the following two situations.
\begin{itemize}
\item[(i)] Smooth class: For a fixed $b$, there exists at least one $f$ in the 2d phase which has semionic statistics with $b$. Then the $m$-string can end on the boundary with endpoint dressed by $f$. 
\item[(ii)] Rough class: For a fixed $b$, no such $f$ is available in the added 2d phase, and the $m$-string cannot end on the boundary. 
\end{itemize}
For general 3d topological phases characterized by the representation category of a finite group $G$, different classes of gapped boundaries correspond to different normal subgroups of $G$, see section \ref{sec:discussion}.

\section{Coupled layer construction}
\label{sec:layer}
In this section, we make use of the coupled layer construction of the 3d bosonic and fermionic toric codes and discuss their gapped boundary theories. 

Consider one stack of (2+1)d Abelian topological orders, each with the K-matrix Chern-Simons Lagrangian \cite{read1990excitation, blok1990effective,wen1992classification},
\be
\L_l = \frac{K_{IJ}}{4\pi} a_{l,\mu}^I \partial_{\nu} a_{l,\lambda}^J.
\ee
where $l\in \{1,2,\cdots, L_0\}$ is the layer index and $I\in \{1,2,\cdots, r\}$ with $r=\text{dim}(K)$. We can condense composite particles in each two consecutive layers to get a 3d topological phase. This procedure was previously described in \cite{ChaoMing}. Denote the linearly-independent composite particles by
\be
n_i^{(l)}=p_i\otimes z_l + q_i\otimes z_{l+1},
\ee
where $p_i$ and $q_i$ are $r$-component integer vectors that label the quasiparticle types in layer $l$ and $l+1$, respectively, and $z_l$ is an $L_0$-component unit vector with the $l$-th entry being one and all the rest being zero. For the condensates to be bosonic, we require
\be
n_i^{(l)T}\mathcal{K}^{-1} n_{j}^{(l')}=0,\quad \forall i, j, l, l'.
\label{eq:null1}
\ee
Here $\mathcal{K}=K\otimes \mathbbm{1}_{L_0\times L_0}$ is an extended $K$ matrix for the stack, and $i\in\{1,2,\cdots r/2\}$. Equation \eqref{eq:null1} can be rewritten explicitly using the data of the 2d layers \footnote{Equation \eqref{eq:null} is not the most general null condition that one can write down, see for example \cite{PhysRevB.88.235103} section IV. B. 2. But this fact is not very important for the purpose of this work. }
\be
p_i^T K^{-1} p_j + q_i^T K^{-1} q_j=0,\ \ p_i^T K^{-1} q_j=0,\ \ \forall i, j.
\label{eq:null}
\ee
The choice of sets $\{p_i\}$ and $\{q_i\}$ is defined modulo a linear transformation by an integer unimodular matrix $U$ that acts as $\bm{p}'=U\bm{p}$ and $\bm{q}'=U\bm{q}.$

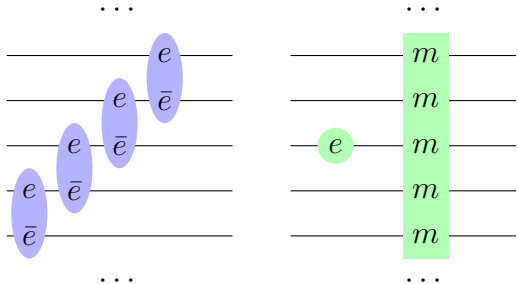
\begin{figure}[htbp]
\centering
\begin{tikzpicture}[scale=0.6]
\draw (0,0)--(5,0);
\draw (0,1)--(5,1);
\draw (0,2)--(5,2);
\draw (0,3)--(5,3);
\draw (0,4)--(5,4);
\fill[blue!30] (0.5,0.5) ellipse (0.4 and 1);
\node at (0.5,0) {{\large $\bar{e}$}};
\node at (0.5,1) {{\large $e$}};
\fill[blue!30] (1.5,1.5) ellipse (0.4 and 1);
\node at (1.5,1) {{\large $\bar{e}$}};
\node at (1.5,2) {{\large $e$}};
\fill[blue!30] (2.5,2.5) ellipse (0.4 and 1);
\node at (2.5,2) {{\large $\bar{e}$}};
\node at (2.5,3) {{\large $e$}};
\fill[blue!30] (3.5,3.5) ellipse (0.4 and 1);
\node at (3.5,3) {{\large $\bar{e}$}};
\node at (3.5,4) {{\large $e$}};
\node at (2.5,-1) {{\large $\cdots$}};
\node at (2.5,5) {{\large $\cdots$}};
\end{tikzpicture}
\quad\quad
\begin{tikzpicture}[scale=0.6]
\draw (0,0)--(5,0);
\draw (0,1)--(5,1);
\draw (0,2)--(5,2);
\draw (0,3)--(5,3);
\draw (0,4)--(5,4);
\fill[green!30] (1,2) ellipse (0.4 and 0.4);
\node at (1,2) {{\large $e$}};
\fill[green!30] (2.5,-0.5) rectangle (3.5,4.5);
\node at (3,-1) {{\large $\cdots$}};
\node at (3,0) {{\large $m$}};
\node at (3,1) {{\large $m$}};
\node at (3,2) {{\large $m$}};
\node at (3,3) {{\large $m$}};
\node at (3,4) {{\large $m$}};
\node at (3,5) {{\large $\cdots$}};
\end{tikzpicture}
\caption{Bosonic exciton condensations give rise to 3d bosonic $\Z_N$ toric code. Black lines are 2d $\Z_N$ toric code layers. Left: Blue nodes describe condensates. Right: Green nodes describe deconfined excitations. }
\label{fig:stack_btc}
\end{figure}

As an example, take conventional $\mathbb{Z}_n$ topological order in each layer, with $K=N \sigma_x$. Consider bosonic exciton/dipole condensation give by
\be
p=(-1,\ 0)^T,\quad q=(1,\ 0)^T, 
\label{eq:basic_condensate}
\ee
where periodic boundary condition is assumed. An intuitive picture can be found in figure \ref{fig:stack_btc}. The charge excitations $e$ in the 2d layers can now combine with the condensates, hop vertically and become a deconfined quasiparticle in 3d. The magnetic string commutes with all condensates and remains deconfined. So we have a 3d bosonic toric code (of type $e_b m_b$).

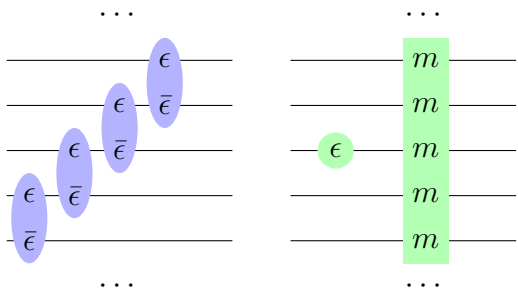
\begin{figure}[htbp]
\centering
\begin{tikzpicture}[scale=0.6]
\draw (0,0)--(5,0);
\draw (0,1)--(5,1);
\draw (0,2)--(5,2);
\draw (0,3)--(5,3);
\draw (0,4)--(5,4);
\fill[blue!30] (0.5,0.5) ellipse (0.4 and 1);
\node at (0.5,0) {{\large $\bar{\epsilon}$}};
\node at (0.5,1) {{\large $\epsilon$}};
\fill[blue!30] (1.5,1.5) ellipse (0.4 and 1);
\node at (1.5,1) {{\large $\bar{\epsilon}$}};
\node at (1.5,2) {{\large $\epsilon$}};
\fill[blue!30] (2.5,2.5) ellipse (0.4 and 1);
\node at (2.5,2) {{\large $\bar{\epsilon}$}};
\node at (2.5,3) {{\large $\epsilon$}};
\fill[blue!30] (3.5,3.5) ellipse (0.4 and 1);
\node at (3.5,3) {{\large $\bar{\epsilon}$}};
\node at (3.5,4) {{\large $\epsilon$}};
\node at (2.5,5) {{\large $\cdots$}};
\node at (2.5,-1) {{\large $\cdots$}};
\end{tikzpicture}
\quad \quad 
\begin{tikzpicture}[scale=0.6]
\draw (0,0)--(5,0);
\draw (0,1)--(5,1);
\draw (0,2)--(5,2);
\draw (0,3)--(5,3);
\draw (0,4)--(5,4);
\fill[green!30] (1,2) ellipse (0.4 and 0.4);
\node at (1,2) {{\large $\epsilon$}};
\fill[green!30] (2.5,-0.5) rectangle (3.5,4.5);
\node at (3,-1) {{\large $\cdots$}};
\node at (3,0) {{\large $m$}};
\node at (3,1) {{\large $m$}};
\node at (3,2) {{\large $m$}};
\node at (3,3) {{\large $m$}};
\node at (3,4) {{\large $m$}};
\node at (3,5) {{\large $\cdots$}};
\end{tikzpicture}
\caption{Fermionic exciton condensations give rise to 3d fermionic $\Z_N$ toric code.}
\label{fig:stack_ftc}
\end{figure}

One can also consider the following fermionic exciton condensation as depicted in figure \ref{fig:stack_ftc}:
\be
p=(-1,\ -1)^T,\quad q=(1,\ 1)^T.
\label{eq:basic_f_condensate}
\ee
Now the fermions $\epsilon=e\times m$ are mobile in three dimensions, while the $m$-string also remains deconfined. We get the fermionic toric code model of type $(e_fm_b)$.

\subsection{3d bTC with boundary}
\label{subsec:stack_bTC}

Now we consider a gapped boundary of the bosonic toric code on the top. The usual smooth boundary corresponds to simply the same left panel of figure \ref{fig:stack_btc}, but without the dots on top. The $e$ particles are confined on the boundary and $m$ string can safely end on the boundary. 

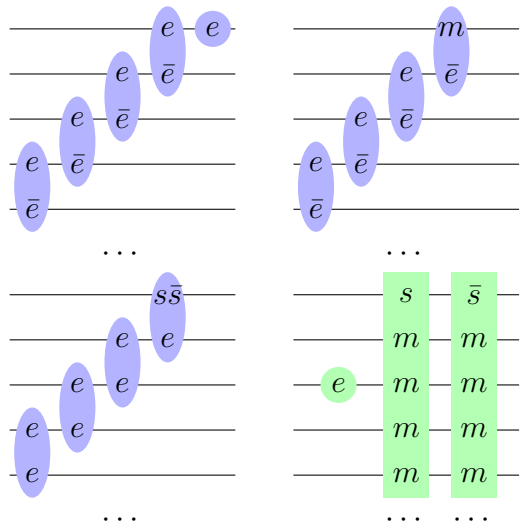
\begin{figure}[htbp]
\centering
\begin{tikzpicture}[scale=0.6]
\draw (0,0)--(5,0);
\draw (0,1)--(5,1);
\draw (0,2)--(5,2);
\draw (0,3)--(5,3);
\draw (0,4)--(5,4);
\fill[blue!30] (0.5,0.5) ellipse (0.4 and 1);
\node at (0.5,0) {{\large $\bar{e}$}};
\node at (0.5,1) {{\large $e$}};
\fill[blue!30] (1.5,1.5) ellipse (0.4 and 1);
\node at (1.5,1) {{\large $\bar{e}$}};
\node at (1.5,2) {{\large $e$}};
\fill[blue!30] (2.5,2.5) ellipse (0.4 and 1);
\node at (2.5,2) {{\large $\bar{e}$}};
\node at (2.5,3) {{\large $e$}};
\fill[blue!30] (3.5,3.5) ellipse (0.4 and 1);
\node at (3.5,3) {{\large $\bar{e}$}};
\node at (3.5,4) {{\large $e$}};
\node at (2.5,-1) {{\large $\cdots$}};
\fill[blue!30] (4.5,4) ellipse (0.4 and 0.4);
\node at (4.5,4) {{\large $e$}};
\end{tikzpicture}
\quad\quad
\begin{tikzpicture}[scale=0.6]
\draw (0,0)--(5,0);
\draw (0,1)--(5,1);
\draw (0,2)--(5,2);
\draw (0,3)--(5,3);
\draw (0,4)--(5,4);
\fill[blue!30] (0.5,0.5) ellipse (0.4 and 1);
\node at (0.5,0) {{\large $\bar{e}$}};
\node at (0.5,1) {{\large $e$}};
\fill[blue!30] (1.5,1.5) ellipse (0.4 and 1);
\node at (1.5,1) {{\large $\bar{e}$}};
\node at (1.5,2) {{\large $e$}};
\fill[blue!30] (2.5,2.5) ellipse (0.4 and 1);
\node at (2.5,2) {{\large $\bar{e}$}};
\node at (2.5,3) {{\large $e$}};
\fill[blue!30] (3.5,3.5) ellipse (0.4 and 1);
\node at (3.5,3) {{\large $\bar{e}$}};
\node at (3.5,4) {{\large $m$}};
\node at (2.5,-1) {{\large $\cdots$}};
\end{tikzpicture}
\quad \quad
\begin{tikzpicture}[scale=0.6]
\draw (0,0)--(5,0);
\draw (0,1)--(5,1);
\draw (0,2)--(5,2);
\draw (0,3)--(5,3);
\draw (0,4)--(5,4);
\fill[blue!30] (0.5,0.5) ellipse (0.4 and 1);
\node at (0.5,0) {{\large $e$}};
\node at (0.5,1) {{\large $e$}};
\fill[blue!30] (1.5,1.5) ellipse (0.4 and 1);
\node at (1.5,1) {{\large $e$}};
\node at (1.5,2) {{\large $e$}};
\fill[blue!30] (2.5,2.5) ellipse (0.4 and 1);
\node at (2.5,2) {{\large $e$}};
\node at (2.5,3) {{\large $e$}};
\fill[blue!30] (3.5,3.5) ellipse (0.4 and 1);
\node at (3.5,3) {{\large $e$}};
\node at (3.5,4) {{\large $s\bar{s}$}};
\node at (2.5,-1) {{\large $\cdots$}};
\end{tikzpicture}
\quad \quad 
\begin{tikzpicture}[scale=0.6]
\draw (0,0)--(5,0);
\draw (0,1)--(5,1);
\draw (0,2)--(5,2);
\draw (0,3)--(5,3);
\draw (0,4)--(5,4);
\fill[green!30] (1,2) ellipse (0.4 and 0.4);
\node at (1,2) {{\large $e$}};
\fill[green!30] (2,-0.5) rectangle (3,4.5);
\node at (2.5,-1) {{\large $\cdots$}};
\node at (2.5,0) {{\large $m$}};
\node at (2.5,1) {{\large $m$}};
\node at (2.5,2) {{\large $m$}};
\node at (2.5,3) {{\large $m$}};
\node at (2.5,4) {{\large $s$}};
\fill[green!30] (3.5,-0.5) rectangle (4.5,4.5);
\node at (4,-1) {{\large $\cdots$}};
\node at (4,0) {{\large $m$}};
\node at (4,1) {{\large $m$}};
\node at (4,2) {{\large $m$}};
\node at (4,3) {{\large $m$}};
\node at (4,4) {{\large $\bar{s}$}};
\end{tikzpicture}
\caption{Left top: rough boundary of 3d bTC. Right top: exchange smooth boundary. Left bottom: condensates for the twisted smooth boundary at $N=2$. Left right: deconfined excitations for the twisted smooth boundary.}
\label{fig:semion}
\end{figure}

The rough boundary corresponds to additionally condensing charge $e$ on the top layer, and the $m$ string can no longer end on the boundary. In the bulk, the condensates are described by the same $p=(-1,\ 0)^T$ and $q=(1,\ 0)^T$ as before when constructing the 3d bTC, but near the boundary, we have another $\tilde{p}=(1,\ 0)^T$ condensed on the top layer. An intuitive picture is shown in the left top panel of figure \ref{fig:semion}.

Another gapped boundary condition exists in the literature for $N=2$\cite{Yidun,Janet}, called the twisted smooth boundary. One can obtain this boundary theory again using the coupled layer construction \cite{Janet}, but now instead of coupling layers of 2d toric codes, on the boundary we add one layer of double semion, with $\tilde{K}=2\sigma_z$. In the bulk the condensates are again the same as \eqref{eq:basic_condensate}. Near the boundary, however, the top two layers have the following condensate,
\be
p=(1,\ 0)^T,\quad \tilde{q}=(1,\ 1)^T.
\ee
The tilde on $q$ indicates that it lives in the double semion layer. This is depicted in the left bottom panel in figure \ref{fig:semion}, and the right bottom panel shows the deconfined excitations. The $e$ particle can freely move in 3d, but upon touching the boundary, it becomes the boson $s\bar{s}$. The $m$-string can still end on the boundary but the endpoint can be either the semion $s$ or the anti-semion $\bar{s}$.  

Yet another possible boundary condition is allowed, by again taking the 2d TO to be $\Z_N$ toric code, but condensing near the boundary the composite of $\bar{e}\otimes m$: 
\be
p=(-1,\ 0)^T,\quad \tilde{q}=(0,\ 1)^T. 
\ee
Since $m$ is a boson, the condensation is allowed. This means the charge $e$ becomes $m$-anyon when hopping onto the boundary, while the $m$-string has its endpoint being an $e$-anyon on the boundary. We name this the \textit{exchange smooth boundary} because it is obtained from the smooth boundary by exchanging $e\leftrightarrow m$ on the boundary. 

One can consider adding a more general $\tilde{K}$-theory on the top layer instead of the double semion model, as long as there exists a  boson $\tilde{q}$ in the theory. Then one can condense the $p=(-1,\ 0)^T$ in the next-to-boundary toric code layer together with this $\tilde{q}$. The $e$-charge in the bulk will become $\tilde{q}$ when reaching boundary, and the $m$-string can end on the boundary if and only if in the added layer, there further exists at least one excitation $\tilde{i}$ such that $\tilde{q}^T \tilde{K}^{-1} \tilde{i} = 1/N$. 

Naïvely there seems to be an infinite number of gapped boundary conditions for the 3d bTC, corresponding to different combinations of choices $\{\tilde{K},\tilde{q}\}$. But they fall naturally into two \textit{classes} based on whether there the $m$-strings can end on the boundary or not, regardless of what lives on the endpoint. The usual smooth and rough boundaries are thus typical representatives of these two classes, respectively.

\subsection{3d fTC with boundary and more general models}
\label{subsec:layer_fTC}
The gapped boundary theory for the fermionic toric code is similar. In the bulk, take the condensate \eqref{eq:basic_f_condensate} as usual, while on the boundary, we can add a top layer with general $\tilde{K}$, as long as it includes some fermionic excitation $\tilde{q}$, such that together with $p=(-1,\ -1)^T$, they satisfy
\be
p_i^T K^{-1} p_j + \tilde{q}_i^T \tilde{K}^{-1} \tilde{q}_j=0,\quad \forall i, j.
\label{eq:coupled_criteria}
\ee

One can be more general and similarly study the gapped boundaries for all 3d theories that can arise from the coupled layer construction using $K$-matrix Chern-Simons theories. If the bulk theory is characterized by the set $\{p_i, q_j\}$ in equation \eqref{eq:null}, on the boundary we condense $\{p_i,\tilde{q}_j\}$ that again satisfy \ref{eq:coupled_criteria}.  One can again discuss whether string-like excitations can end on the boundary by looking at its mutual statistics with the condensate.

\section{Lattice model}
\label{sec:WW}
In this section, we provide a microscopic description for the discussions in section \ref{subsec:stack_bTC}. We will use a Walker-Wang model \cite{WW} in the bulk, which describes a large class of (3+1)d topological phases, and a string net model \cite{SN} on the surface, which describes a large class of (2+1)d topological phases,  and then couple them together. 

\subsection{Walker-Wang model in the bulk}

\begin{figure}[htbp]
\centering
\includegraphics[scale=0.25]{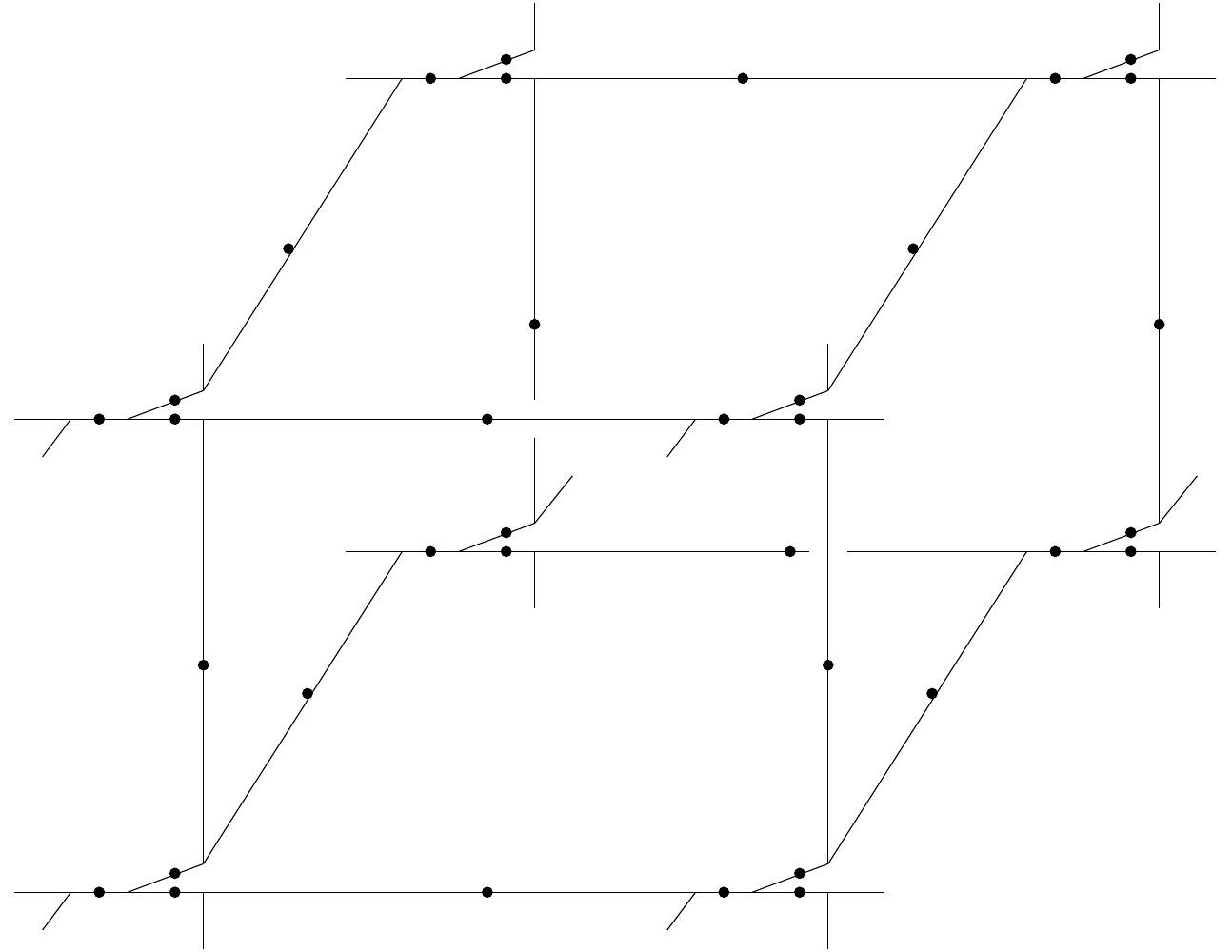}
\caption{Left: Walker-Wang trivalent graph.}
\label{fig:WW}
\end{figure}

The walker-Wang geometry is shown in figure \ref{fig:WW}.
For convenience, in this section we will focus on the $\Z_2$ case in the bulk. Extensions to the $\Z_N$ cases can be straightforwardly accomplished by using generalized Pauli matrices and keeping track of the hermitian conjugates correctly. On each edge of fig. \ref{fig:WW}, there lives a spin-1/2 degree of freedom. The Walker-Wang Hamiltonian takes the form \cite{WW}
\be
\begin{split}
& H_{\text{WW}}=-\sum_v A_v -\sum_p B_p,\\
& A_v=\prod_{i\in s(v)} Z_i,\quad B_p= \prod_{i\in\partial p}X_i,
\end{split}
\label{eq:WW}
\ee
where $s(v)$ is the set of three edges attached to vertex $v$ and $\partial p$ is the set of ten edges of a plaquette $p$. All the terms in the Hamiltonian mutually commute. The ground state satisfies $A_v=1=B_p$ for all $v$ and $p$, and the ground state degeneracy is $2^3$ on $T^3$. 

There are two types of excitations corresponding to the violations of the vertex and plaquette terms, respectively. The excitation creation operators are
\be
W_e(\C) = \prod_{i\in \C} X_i, \quad W_m(\S) = \prod_{i\in\S} Z_i,
\ee
where $\C$ is a 1d path on the lattice. $\S$ is a 2d membrane on the dual lattice and $Z_i$'s act on the edges piercing the membrane. $W_e(\C)$ commutes with the Hamiltonian except at its endpoints, so the defects on the endpoints are deconfined. $W_m(\S)$ has an energy cost which scales linearly with the length of $\partial \S$.

\subsection{String-net on the boundary}

\begin{figure}[htbp]
\centering
\includegraphics[scale=0.17,trim=0.1cm 0 0 0.07cm ,clip]{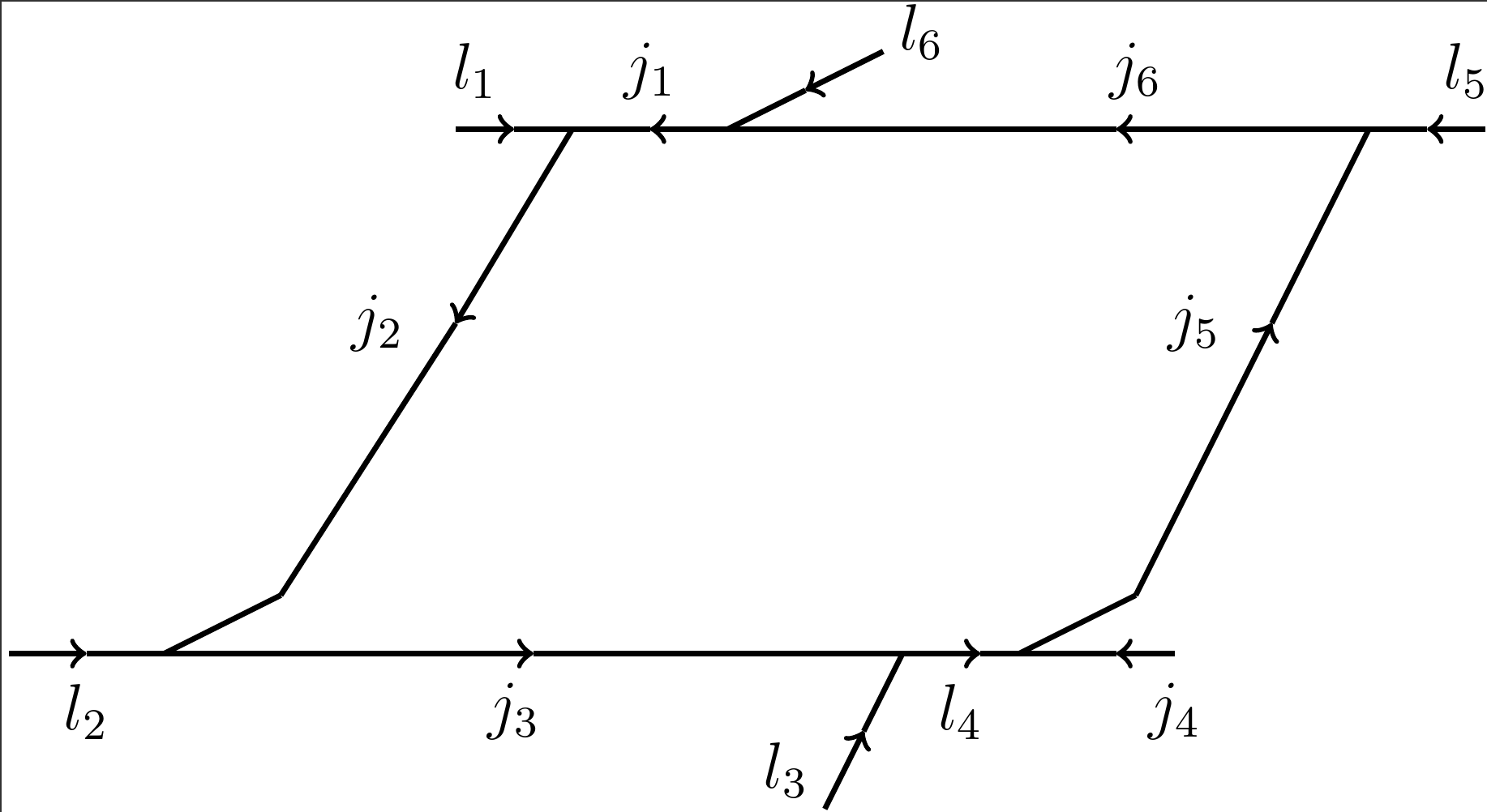}\\
\ \\
\includegraphics[scale=0.3]{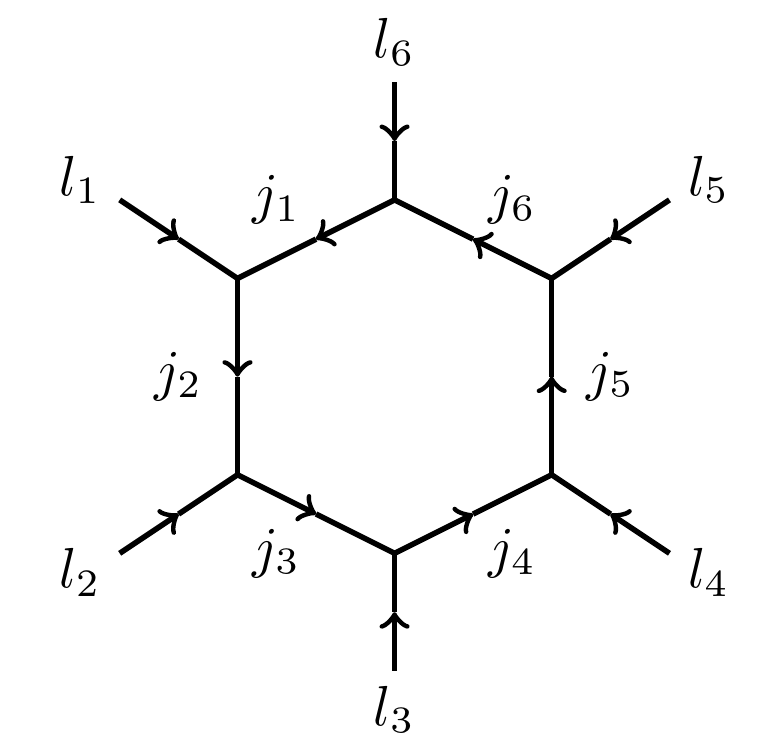}
\caption{Surface geometry of Walker-Wang (top), which is nothing but the honeycomb lattice (bottom).}
\label{fig:surface}
\end{figure}

\begin{figure}[htbp]
    \centering
    \includegraphics[scale=0.17]{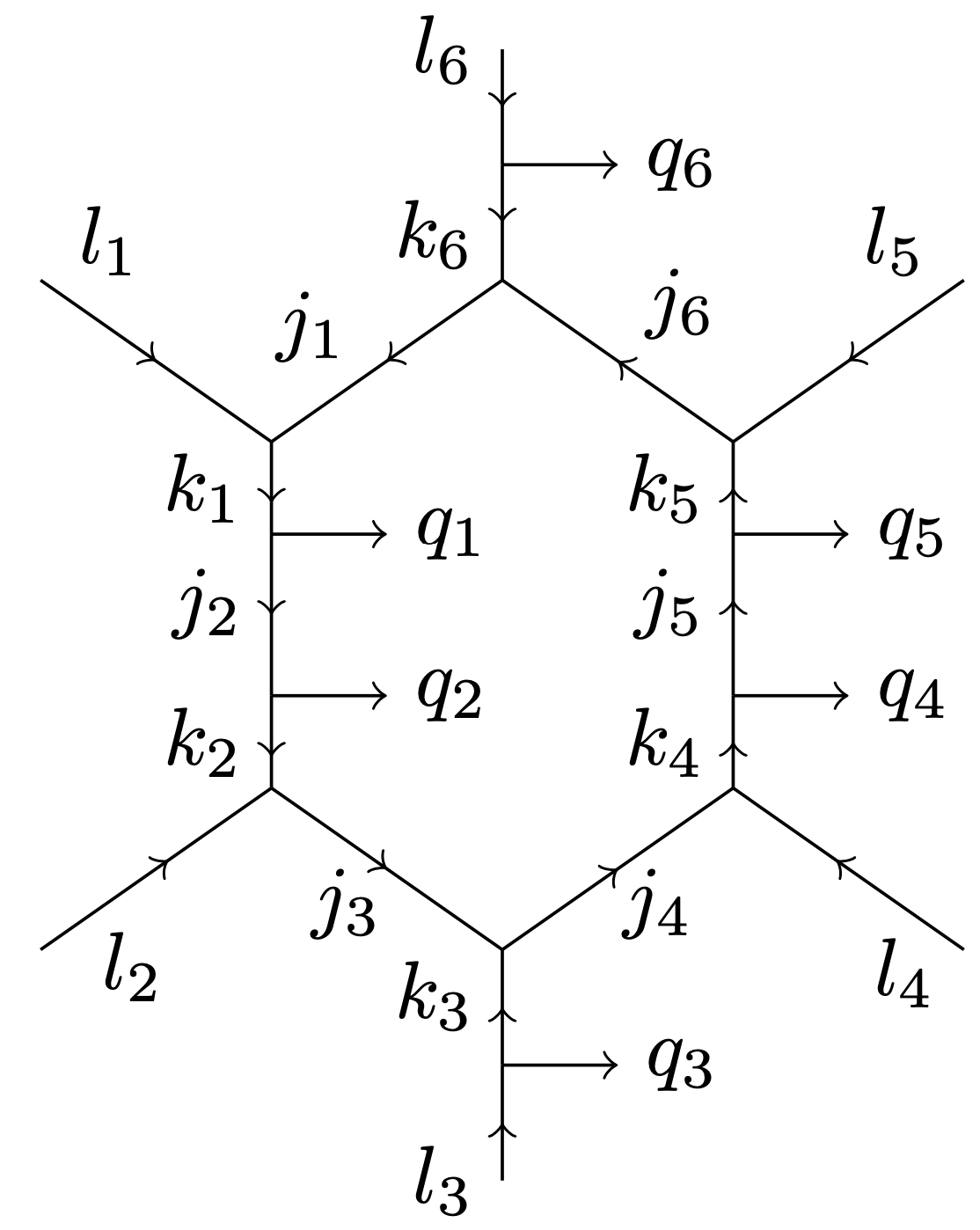}
    \caption{Short charge tails $q_i$'s are added near each vertex.}
    \label{fig:extended}
\end{figure}

We define the string-net model \cite{SN} on a lattice that looks like the top surface of the Walker-Wang model, where all the open tails pointing out of the surface are removed, see figure \ref{fig:surface}. This can be easily deformed into a honeycomb lattice, on which we define our string-net model. 
The basic data to define a string-net model includes $I=\{0,1,\cdots, N-1\}$ a set of labels assigned to the directed links, $d_i\in \mathbb{R}$ quantum dimension of each label $i\in I$, $N_{ijk}\in \mathbb{Z}_0^+$ the fusion rule for $i, j, k\in I$ which describes how different links should meet at a vertex, and (symmetrized) 6j-symbols $G_{ijk}^{lmn}\in \mathbb{C}$. In addition, each label $j$ has a conjugate $j^*\in I$, and taking the conjugate amounts to reversing the direction (arrow) of the link. In the original string-net model, it is easy to describe single flux excitations but hard to describe dyonic and many-flux excitations. So here we will make use of the extended string-net model developed in \cite{Full}, and many examples therein. The extension simply works as follows: we associate to each vertex a short open tail $q_i$, such that the bottom panel of \ref{fig:surface} turns into \ref{fig:extended}. 
For simplicity, from now on we restrict to the multiplicity-free case where $N_{ijk}=\delta_{ijk}\in\{0,1\}$. The Hilbert space is spanned by different labelings of all the links on the honeycomb lattice that satisfy the hard constraint $\delta_{ijk}=0$ near each vertex. The extended string-net Hamiltonian consists of two terms,
\be
H_{\text{SN}}=-\sum_v \Q_v - \sum_p \B_p, \quad \B_p=\frac{1}{D}\sum_s d_s \B_p^s,
\label{eq:SN}
\ee
where $D=\sum_i d_i^2$ is the total quantum dimension. The $\Q_v$ term imposes the zero-charge constraint near the vertex $v$, 
\be 
\Q_v\vcenter{\hbox{\includegraphics[scale = 0.18]{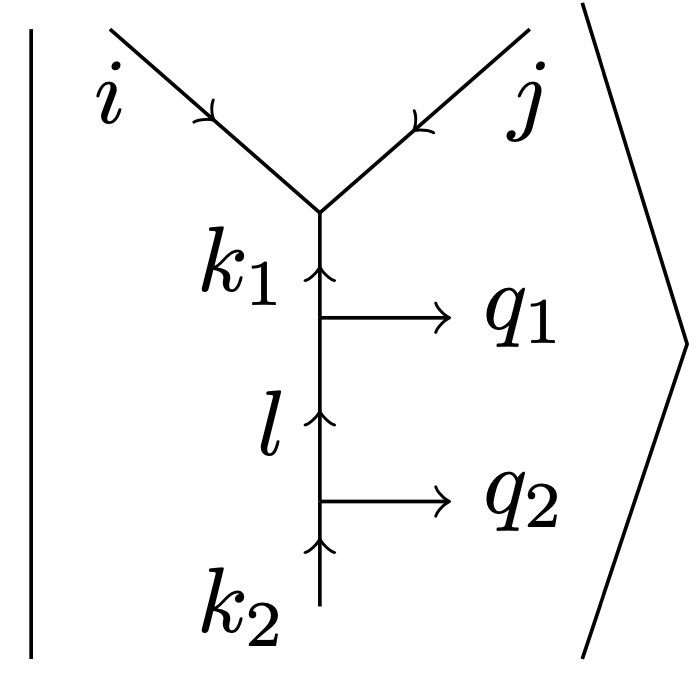}}}=\delta_{q_1,0}\vcenter{\hbox{\includegraphics[scale = 0.18]{fig_Qv.png}}}.
\label{eq:Qv}
\ee
\begin{widetext}
The $\mathcal{B}_p^s$ term acts on plaquette $p$ in the following way:
\be
\begin{split}
    & \mathcal{B}_p^s\  \vcenter{\hbox{\includegraphics[scale = 0.16]{fig_Bp1}}} = \delta_{q_1,0}\delta_{q_2,0}\sum_{j_1',j_2',j_3',j_4',j_5',j_6',k_4',k_5'} \gamma\  \vcenter{\hbox{\includegraphics[scale = 0.16]{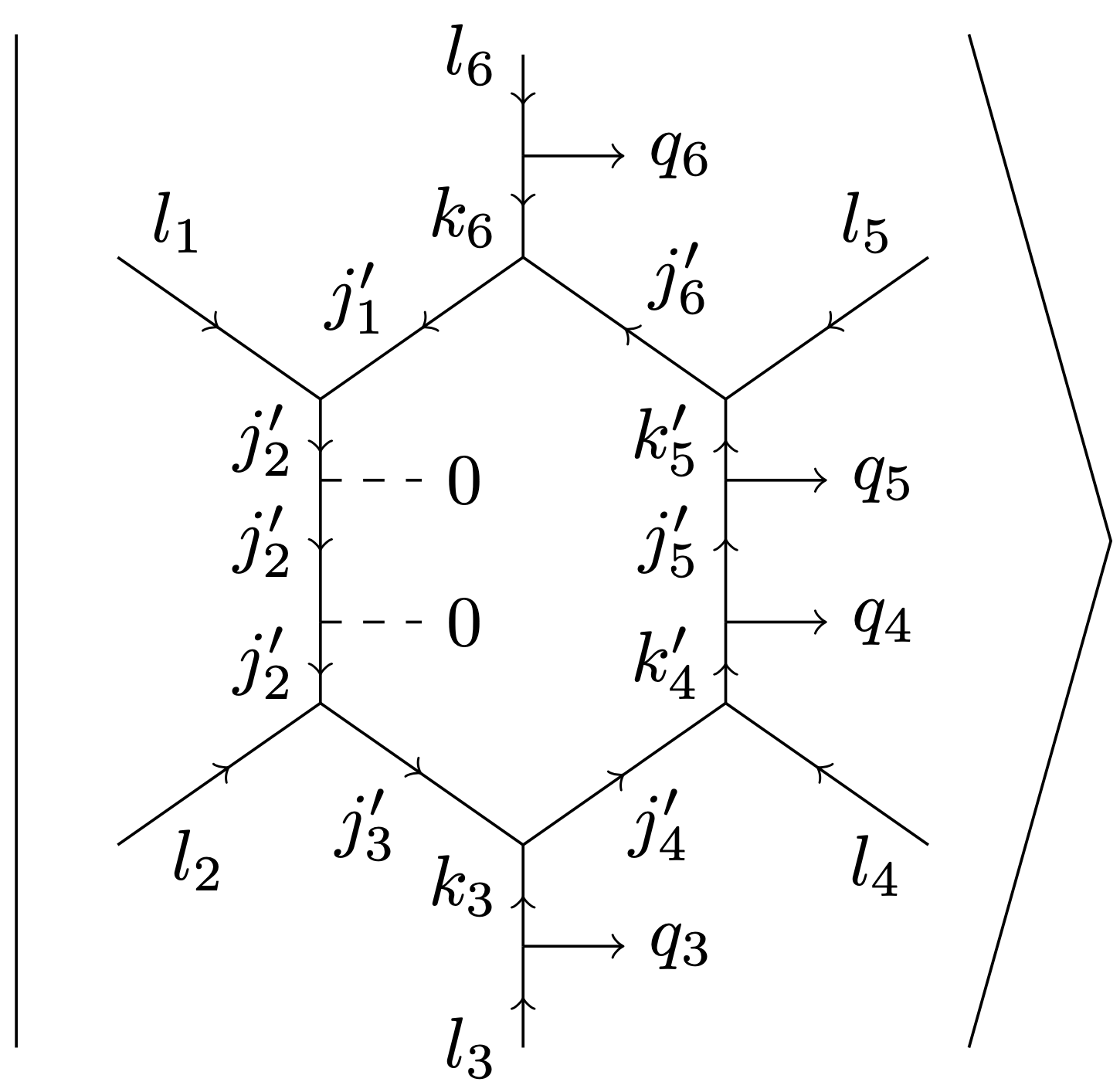}}},\\
    & \gamma\equiv \sqrt{d_{k_4}d_{k_5}d_{k_4'}d_{k_5'}}\big(\prod_{a=1}^6 \sqrt{d_{j_a}d_{j_a'}}\big) G_{s^*j_1'j_2'}^{l_1k_1^*j_1} G_{s^*j_2'j_3'}^{l_2j_3^*k_2}G_{s^*j_3'j_4'}^{k_3j_4^*j_3} G_{s^*j_4'k_4'^*}^{l_4k_4^*j_4}G_{s^*k_5'j_6'^*}^{l_5j_6^*k_5}G_{s^*j_6'j_1'^*}^{k_6j_1^*j_6}G_{s^*k_4'j_5'^*}^{q_4j_5^*k_4}G_{s^*j_5'k_5'^*}^{q_5k_5^*j_5}.
\end{split}
\ee
\end{widetext}
Restriction to the $\Q=1$ subspace recovers the original string-net model. 

Each elementary type of dyonic excitation $J$ in the extended model is associated with an irreducible representation of the tube algebra, and corresponds to a solution for the half-braiding tensor $z_{pjqt}$ in the following equation:
\be
\begin{split}
& \sum_{lrs} d_r d_s z_{lnqr} z_{pmls} G_{nr^*t}^{m^*sl^*}G_{jn^*t}^{s^*pm}G_{q^*n^*k}^{m^*tr^*}\\
=\ & \delta_{mnj^*}\delta_{jk}z_{pjqt}/d_j.
\end{split}
\ee
Lastly we define the dyon creation operator near edge $e$:
\be
\begin{split}
& W_e^{J;pq^*}\vcenter{\hbox{\includegraphics[scale = 0.2]{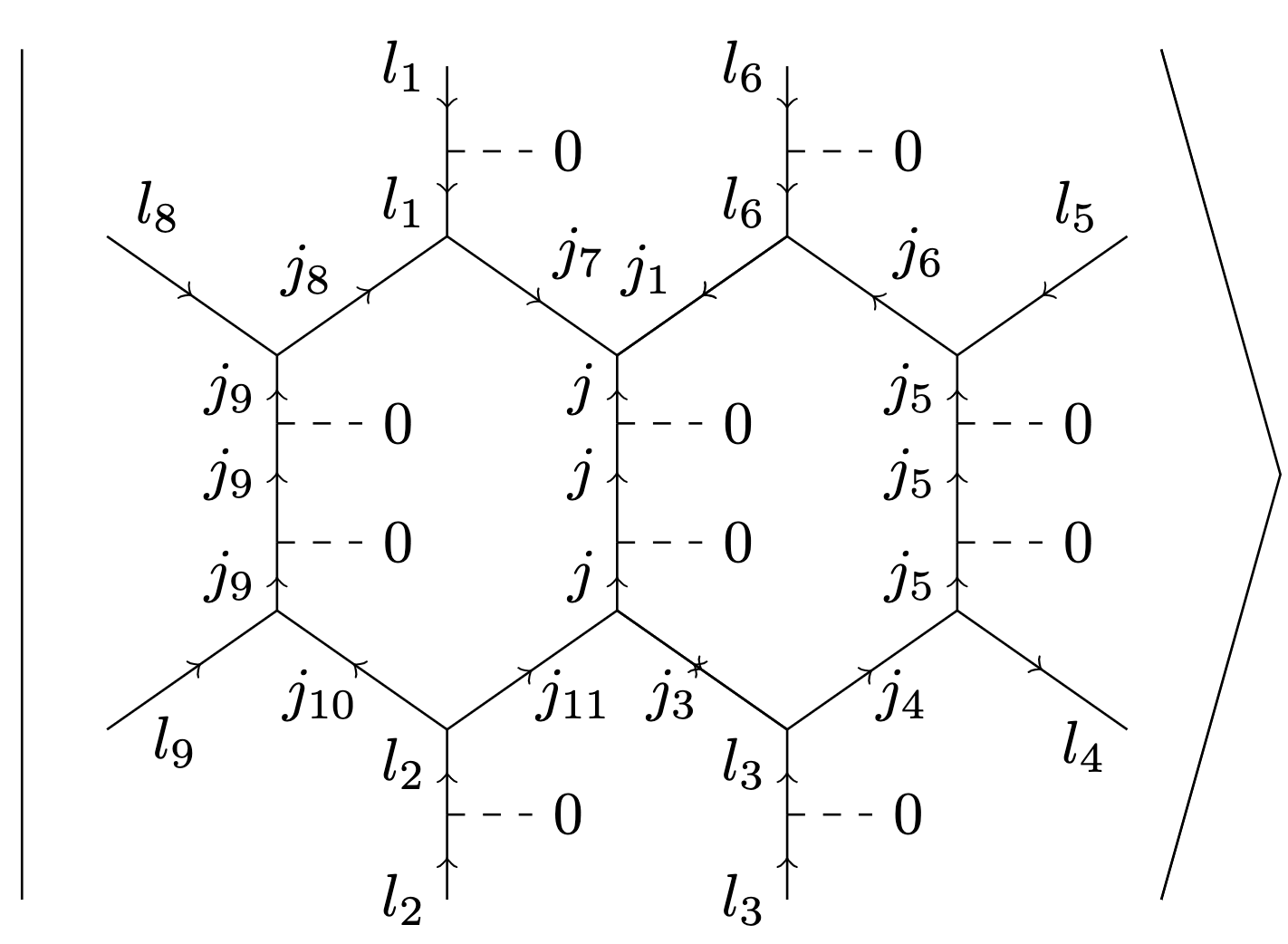}}}\\
= & \sum_{j'}\sqrt{\frac{d_j'}{d_j}}\ \overline{z^J_{pj'qj}}\vcenter{\hbox{\includegraphics[scale = 0.2]{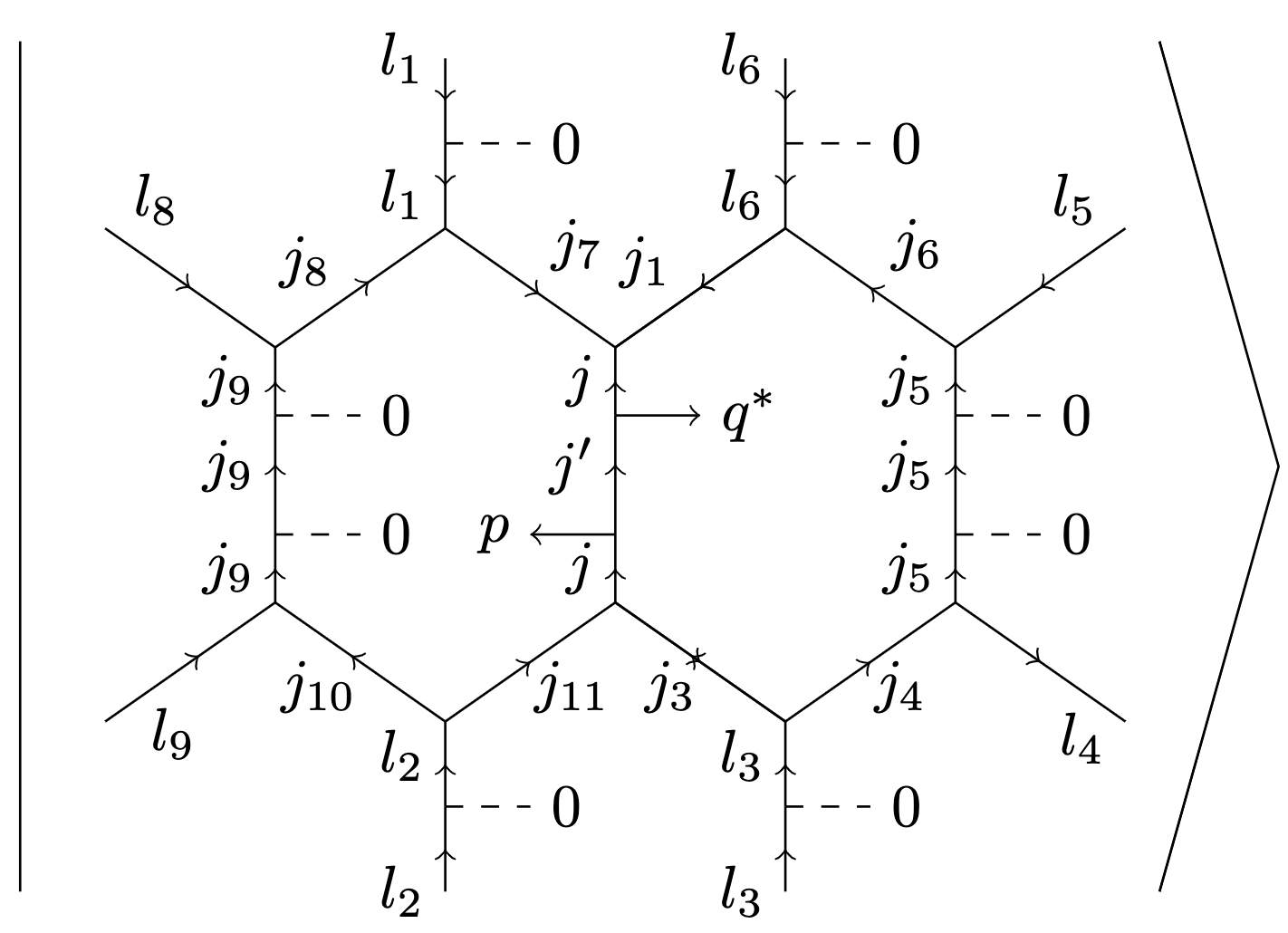}}}
\end{split}
\label{eq:WJ}
\ee
Here charges $p$ and $q^*$ are created at the endpoints of this short dyonic string.

\subsection{Coupling the two models}

We now stack the 2d string-net surface on the top boundary of the Walker-Wang model, such that each link of the honeycomb lattice hosts a $2\times N$-dimensional Hilbert space, 2 coming from the spin-1/2 of Walker-Wang, and $N$ is the cardinality of the label set $I$ of the string-net. 

We couple the $\Z_2$ Walker-Wang and a general string-net model in the following way, 
\be
H_c^J=- \sum_j \sum_{p,q} X_j \otimes W_j^{J;pq^*}+\text{h.c.},
\label{eq:H_c}
\ee
where $j$ is a link on the honeycomb lattice. $X_j$ acts on the Walker-Wang qubit, while 
$W_j^{J,pq^*}$ acts on the string-net link. 
The term $H_c^J$ associates a short open string of charges $e$ in the Walker-Wang model with a short string of bosonic dyon $J$ with charged endpoints $p$ and $q^*$ in the string-net model. $H_c^J$ create, annihilate, and more generally proliferate the composite of $e\otimes J$. As a result, excitations that braid nontrivially with $e\otimes J$ become confined. In addition, deconfined excitations related by fusion with $e\otimes J$ become identified. 

We now fix the other boundary terms. The $A_v$ terms which entirely live on the surface of the Walker-Wang model anti-commute with $H_c^J$. This is natural as such $A_v$ terms create short loops of $m$ which are confined by the $e\otimes J$ condensate. However, if there exists an anyon $K$ in the string-net model which has mutual semionic statistics with $J$, the composite $m\otimes K$ is not confined by the condensate. Therefore, we should include the terms which create short loops of $m\otimes K.$ They read 
\be
\tilde{A}_v^K = \big(\prod_{j\in s(v) }  Z_j\big) \otimes O_v^K, 
\label{eq:modified_A}
\ee
where $O_v^K$ creates a shortest loop of $K$ dyon around the vertex v.  Schematically $O_v^K=\sum_{q_1,q_2,q_3} \mathcal{P} W_{e_3}^{K;q_1^*q_3}W_{e_2}^{K;q_2q_3^*}W_{e_1}^{K;q_1q_2^*}$. The three $W^K$ operators, as defined as in equation \eqref{eq:WJ}, create three shortest open-strings near the three edges surrounding the vertex $v$.  $\mathcal{P}$ contracts the charges which live at the endpoints of these short strings, thus connecting the three pieces into one short closed loop. More details about the contraction of charges are reviewed in appendix \ref{app:contraction}. There is one $\tilde{A}_v^K$ term for each $K$ that has mutual semionic statistics with $J$. The mutual statistics can be derived from the half-braiding tensor in the following way. The modular $S$-matrix can be written as
 \be
 S_{JK}=\sum_{p,q,t} d_t\ \overline{z^{J}_{pqpt} z^{K}_{qpqt}},
 \ee
which is independent of the charge tails $p', q'$ of the short dyon string as long as these charges are allowed to live on the endpoints of the dyon-string. The monodromy matrix, which plays an important role in anyon interferometry \cite{PhysRevLett.98.070401,PhysRevLett.97.016401,BONDERSON20082709} is related to the modular $S$ matrix by
\be
M_{JK}=\frac{S_{JK}^* S_{00}}{S_{0J} S_{0K}}. 
\ee
Mutual semionic statistics between $J$ and $K$ simply means that $M_{JK}=-1.$ 
This requirement guarantees that $O_v^K$ anticommutes with the $W_j^{J}$ in  $H_c^J$, such that $[\tilde{A}_v^K,  H_c^J]=0$. This term can be interpreted as dressing the endpoint of a $m$-string extended from the bulk a dyon $K$. 

We move on to the remaining Hamiltonian terms on the boundary. Deep inside the bulk of the Walker-Wang model, the original stabilizers of \eqref{eq:WW} trivially commute with $H_c^J$ because they have no shared links with $H_c^J$. Near the surface, the $B_p$ terms always commute with $H_c^J$. There are some $A_v$ terms that have two links living on the surface and one link living in the bulk. Such terms also commute with $H_c^J$ because those two links living on the surface of the Walker-Wang model are identified as one link on the honeycomb lattice, see figure \ref{fig:surface}, and acting Pauli matrix $Z$ twice on that degree of freedom is identity. 

Within the string-net model,  
there can exist other dyons $M$ which have trivial statistics with $J$ and are therefore deconfined. They correspond to 2d mobile quasiparticles and are not attached to strings extending into the bulk. We should therefore also include the short loops of such dyons, finally arriving at the full Hamiltonian is $H = H_{\text{bulk},WW} + H_{\text{surface}}$, with 
\be
\begin{split}
 H_{\text{surface}} = & - \sum_{j,p,q}  X_j \otimes W_j^{J;pq^*}- \sum_v \sum_K {}^{'} \tilde{A}_v^K \\
& \quad \ -\sum_p B_p -\sum_v \sum_{M} {}^{'} O_v^{M} + \text{h.c.}
\end{split}
\label{eq:WW_SN_Ham}
\ee
Here the $\sum'_K$ indicates that we only sum over $K$'s that have mutual semionic statistics with $J$, and the $\sum'_M$ means the summation is over $M$'s that has trivial statistics with $J$.

While there are many gapped boundaries of $\Z_2$ Walker-Wang model characterized by a choice of string-net model together with a boson quasiparticle $J$, there are only two \textit{classes} of gapped boundaries corresponding to whether there exists at least one $K$ that is mutual semionic with $J$, i.e., whether there exists at least one $\tilde{A}_v^K$ term.

\subsubsection{Example: rough boundary}
In the rough boundary case, the string-net model is taken to be trivial, i.e. $I=\{0\}$, $d_0=1$ and $G_{000}^{000}=1,$ $z_{0000}^{J=0}=1.$ The $W^J$ operator is thus trivial and 
\be
H_c^J=-\sum_j X_j\otimes \mathbbm{1}.
\ee
Not surprisingly, this is simply condensing the charges in the toric code on the surface. Now there does not exist any dyon $K$ that has mutual semionic statistics with the $J=0$, so there is no $\tilde{A}_v^K$ term and the m-string cannot end on the boundary. Nor is there a nontrivial $M$ that trivially braid with $J$. So the surface Hamiltonian only contains $H_c^J$ and we get the expected rough boundary.

\subsubsection{Examples: (exchange) smooth boundary}

In the cases of  smooth boundary and exchange smooth boundary (the latter is defined in section \ref{subsec:stack_bTC}), the string-net is taken to be the (2+1)d toric code model. The string-net data for $\Z_{\tilde{N}}$ toric code models are\footnote{There is a typo in ref. \cite{Full} regarding half-braiding tensor of $\Z_{\tilde{N}}$ model. The equation in the current paper is correct.}
\be
\begin{split}
& I=\{0,1,\cdots, \tilde{N}-1\},\quad d_i=1\ \forall i\in I,\\ & N_{ijk}=\delta_{ijk}=1\ \text{iff}\ i+j+k=0 \mod \tilde{N}. \\
& G_{kln}^{ijm}=\delta_{ijm}\delta_{klm^*}\delta_{jkn^*}\delta_{inl},\\  & z^{(g,\mu)}_{pjqt}=\delta_{p,\mu}\delta_{q,\mu}e^{2\pi \mathrm{i} j g/N} \delta_{pjt^*}\delta_{jqt^*}.\\
\end{split}
\ee
Notice that the dyons are labeled by $(g,\mu)$ where $g\in I$ labels the flux and $\mu\in I$ labels the charge. We will take $\tilde{N}=2$ for convenience, the general $\tilde{N}$ case is a straightforward generalization. 

For the smooth boundary, we take the $J=(0,1)$ in $H_c^J,$ such that $W^{J;pq^*}_{j}$, when acting on a graph as in \eqref{eq:WJ}, is only nonzero when 
$p, q=1$, and the amplitude is $ \overline{z^J_{1 j' 1 j}}=(1-\delta_{jj'}).$ 
Consequently the $H_c^J$ associates a charge of Walker-Wang to a charge of string-net. The $K=(1,0)$ dyon has mutual semionic statistics with $J$, and $W^{K;pq^*}$ is nonzero when $p, q=0$, and the amplitude in eqn. \eqref{eq:WJ} is simply $\overline{z^J_{0 j' 0 j}}=\delta_{jj'} e^{-\mathrm{i} \pi j}.$ One can easily double check that 
$W^J_{e}W^K_{e}=-W^K_{e}W^J_{e}$, such that the $\tilde{A}_v^K$ term commutes with $H_c^J$. There is nothing left in the string-net model that is both nontrivial and trivially commutes with $W^J$ and $W^K$. So there is no $W^M$ term in $H_{\text{surface}}.$ 

For the exchange smooth boundary, we take $J=(1,0)$ in $H_c^J$, which associates a charge of Walker-Wang to a flux of string-net, and $K=(0,1)$ is the endpoint of the $m$-string.

\subsubsection{Example: twisted smooth boundary}
Choose the string-net model to be that of the double semion, with the following data 
\be
\begin{split}
& I=\{0,1\},\quad d_0=1,\quad d_1=-1,\\ 
& N_{110}=N_{101}=N_{011}=1,\\
& G_{000}^{000}=1,\quad G_{111}^{000}=i,\quad G_{011}^{011}=-1,\quad R_{11}^0=\mathrm{i},\\
& z^{i\bar{j}}_{pjqt}=\sum_{a,b}d_a d_b R^{a}_{ik} R^{b}_{jk}G^{a^* ik}_{bj^*t} G^{ij q^*}_{t^*ka^*} G^{ibt^*}_{k^*p^*j^*}.
\end{split}
\ee
The dyon excitations are labeled by $i\bar{j}$ with $i,j\in I.$ The twisted smooth boundary is then obtained from taking $J=1\bar{1}$ and $K_1=1\bar{0}$ as well as $K_2=0\bar{1}$ are both allowed. 

\subsubsection{Example: non-abelian surface anyons} 
\label{subsec:Ising}

We now describe a simple example where the surface anyons are non-abelian. We choose the string-net to be that of the doubled Ising model. The label set is $I={0,1,2}$, or ${1,\sigma,\psi}$. The quantum dimensions and fusion rules are
\be
\begin{split}
& d_0=d_2=1,\quad d_1=\sqrt{2},\\
& N_{000}=N_{011}=N_{022}=N_{112}=1.
\end{split}
\ee
The independent 6j-symbols are
\be
\begin{split}
& G_{000}^{000}=G_{222}^{000}=G_{022}^{022}=1,\ \\
& G_{011}^{011}=G_{211}^{011}=-G_{112}^{112}=2^{-1/2},\\
& G_{111}^{000}=G_{122}^{011}=2^{-1/4}. 
\end{split}
\ee
The anyons are again labeled by $i\bar{j}$ with $i, j\in I$. There are three bosons $0\bar{0}$, $1\bar{1}$ and $2\bar{2}$. Choosing for example $J=2\bar{2}$ in equation \eqref{eq:WW_SN_Ham}, the charge $e$ in the toric code becomes $2\bar{2}$ or $\psi\bar{\psi}$ upon reaching the boundary. The $m$-string is allowed to end on the boundary, with possible endpoints $K\in \{0\bar{1},1\bar{0},1\bar{2},2\bar{0}\}$. All these endpoints are non-abelian.

\section{3d Field theory description}
\label{sec:field}
The field theory for the (3+1)d bosonic toric code is
\be
\L=\frac{N}{2\pi} b da,
\ee
with $a$ a one-form field and $b$ a two-form field. Without any boundaries, the generators of the one- and two-form symmetries are
\be
W_m(\S)=e^{i\int_{\S} b},\quad W_e=e^{i\int_{\mathcal{C}} a}.
\label{eq:TC_wilson}
\ee
Here $\mathcal{S}$ is a closed 2d spatial membrane and $\mathcal{C}$ a closed spatial 1d path. They satisfy the following $\Z_N$ Heisenberg algebra
\be
W_m(\S)W_e(\C)=e^{2\pi i I(\S,\C)/N} W_e(\C)W_m(\S),
\ee
where $I(\S,\C)$ is the intersection number between $\S$ and $\C$. 

When there is a boundary at $z=0$, the variation of action contains two parts: one imposes the usual equations of motion in the bulk, and the other is a boundary piece 
\be
\delta S \mid = \frac{N}{2\pi} \int_{z=0} d^3x\    b\delta a
\label{eq:deltaS_bdry}
\ee
To recover the bulk equation of motion, the boundary piece must vanish. Below we will examine two strategies to make $\delta S|=0$, one is to directly choose certain gauge fields to vanish at the boundary, the other is a longer path: first choose the ``temporal gauge'' and introduce additional boundary dynamical fields to replace the gauge fields, such that the bulk gauge symmetries will become global symmetries on the boundary; then further add potentials  to gap out the boundary. The two strategies will lead to equivalent results, but the latter is more convenient to study anomaly inflow, which we will discuss in subsection \ref{subsec:inflow}. 

\subsection{The direct path}

Two obvious gapped boundary conditions that can make \eqref{eq:deltaS_bdry} vanish would be $a\mid =0$ for rough boundary, and $b\mid=0$ for smooth boundary. 

One can also obtain the twisted smooth boundary by adding another (2+1)d TO on the boundary. The TO can be either a chiral Chern-Simons theory or a double semion model. 
The (2+1)d chiral Chern-Simons theory at level $N$ is
\be
\tilde{\L} = \frac{N}{4\pi} ada. 
\label{eq:chiral}
\ee
With this additional term, the variation of the full system on the boundary, including that of the original toric code and the chiral Chern-Simons term, is
\be
\delta S_{\text{tot}}| = \frac{N}{2\pi} \int_{z=0} d^3x\ (\delta a)(b+da).
\label{eq:var_S_tot}
\ee
There is thus an obvious boundary condition of $(b+ da)|=0$. The membrane $\mathcal{S}$ of the Wilson operator $W_m(\S)$, when touching the boundary, gets modified to 
\be
\tilde{W}_m(\S)=e^{i\int_{\S} (b+da)}.
\ee
The modified magnetic Wilson operators $\tilde{W}_m(\S)$ now have nontrivial commutation relations among themselves:
\be
\tilde{W}_m(\S) \tilde{W}_m(\S')= e^{2\pi i I(\S,\S')/N} \tilde{W}_m(\S') \tilde{W}_m(\S).
\ee
When $N=2$, we recover the semionic behavior for the endpoints of the $m$-strings. 

Focusing on $N=2$, we now add the double semion model instead. We write it in a basis such that the $s\bar{s}$ anyon is a charge for a single gauge field,
\be
\tilde{\L}=\frac{K_{IJ}}{4\pi}A_I dA_J,\quad K=\left(\begin{matrix} 0 & 2 \\ 2 & 2 \end{matrix}\right).
\label{eq:ds}
\ee
This can be obtained from the more familiar form of $K'=2\sigma_z$ through the transformation $K=W^T K' W$ with $W=\left(\begin{matrix} 1 & 1 \\ 1 & 0 \end{matrix}\right)$. Next, we identify $A_2\equiv a|$. The total variation of full action on the boundary is now
\be
\delta S_{\text{tot}}\mid =\frac{2}{2\pi}\int\limits_{z=0} d^3x \left[ (\delta a) (b+ dA_1+da) +  \delta A_1 da\right].
\ee
The last term just imposes the usual equation of motion $da=0$ in the double semion model. Plugging it into the first term, the problem reduces to the case of adding a chiral Chern-Simons theory \eqref{eq:var_S_tot}. The gapping condition is $(b+dA_1)\mid=0,$ giving the twisted smooth boundary. 

In general, one can add one-form fields $A_I$ and two-form fields $B_J$ with various couplings to the toric code gauge fields on the (2+1)d surface. The newly added Lagrangian could, for example, include the following types of terms $A_Ida$, $B_Ia,$ $bA_I,$ $A_IdA_J,$ $A_IA_JA_K,$ $B_IA_J.$ 
One can also consider adding higher-rank tensor gauge fields or foliated gauge fields. We will present one such example later in section \ref{sec:EBL}.

\subsection{The longer path}
To make the variation of the toric code action on the boundary, equation \eqref{eq:deltaS_bdry}, to vanish, one can alternatively choose following boundary conditions
\be
a_t\mid= 0,\quad b_{tx}\mid=b_{ty}\mid=b_{tz}\mid=0.
\ee
$b_{tz}\mid=0$ is not required but taken for later convenience. We can extend this boundary condition into the bulk as a gauge fixing. This leads to the following equations of motion 
\be
\epsilon_{ijk}\p_i b_{jk}=0,\quad \epsilon_{ij}\p_{i} a_{j}=0.
\ee
The equations of motion can be solved by
\be
b_{ij}=\p_i\vphi_j-\p_j\vphi_i,\quad a_i=\p_i \hat{\vphi},
\label{eq:sol}
\ee
where $\hat{\vphi}$ is a compact zero-form field and $\vphi$ is a compact one-form field. Plugging in, the bulk Lagrangian reduces to a boundary piece
\be
S=\frac{N}{2\pi} \int_{z=0} d^3x\ (\epsilon_{ij}\p_i\vphi_j)\p_0\hat{\vphi}.
\label{eq:boundary_phi}
\ee
In general there velocity terms such as $(\epsilon_{ij}\p_i \vphi_j)^2$ and $(\p_i \hat{\vphi})^2$ are also allowed on the boundary, but they are not important for our purpose. 
The canonical commutation relation from \eqref{eq:boundary_phi} is
\be
[\hat{\vphi}(t,\bx),\epsilon_{ij}\p_i\vphi_j(t,\bx')]=\frac{2\pi i}{N}\delta^{(2)}(\bx-\bx').
\ee
There are a one-form symmetry and a zero-form symmetry on the boundary,
\be
\vphi_j\rightarrow \vphi_j+\alpha_j(\bx),\quad \hat{\vphi}\rightarrow \hat{\vphi}+\hat{\alpha}(\bx).
\label{eq:phi_sym}
\ee
The corresponding Noether's currents are
\be
\begin{split}
& J_0=\frac{1}{2\pi} \epsilon_{ij} \p_i\vphi_j,\quad J_i=0;\\
& \hat{J}_{0j}=\frac{1}{2\pi}\epsilon_{ij} \p_i\hat{\vphi},\quad \hat{J}_{ij}=0;
\end{split}
\label{eq:phi_current}
\ee
with $\p_{\mu} J_\mu=0$ and $\p_{\mu} \hat{J}_{\mu j}=0$.  The corresponding charges are quantized,
\be
\int dx dy\ J_0=n_0,\ \int dx_i\  \hat{J}_{0i}=\hat{n}_i,\ \ n_0, \hat{n}_i\in \mathbb{Z}. 
\label{eq:charge_TC}
\ee
One can further define the vertex operators
\be
\hat{V}=e^{i\hat{\vphi}},\quad V_j=e^{i\oint dx_j \vphi_j},
\ee
$\hat{V}$, $V_j$ and their hermitian conjugates create/annihilate the flux and charge excitations on the surface, which can be seen from their commutation relations with the density operators,
\be
\begin{split}
& [J_0(\bx),\hat{V}(\bx')]=\frac{1}{N}\delta^{(2)}(\bx-\bx')\hat{V}(\bx),\\
& [\hat{J}_{0j}(\bx),V_j(x_i')]=\frac{1}{N}\delta^{(1)}(x_i-x_i')V_j(x_i').
\end{split}
\ee
In addition, they have the expected nontrivial mutual statistics, 
\be
\hat{V}(\bx)V_j(x_i') =e^{i \pi  \text{sgn}(x_i-x_i')/N }V_j(x_i')\hat{V}(\bx).
\ee
The smooth and rough boundaries correspond to adding to the boundary potential terms $-g \cos(N\vphi_1)-g\cos (N\vphi_2)$ or $-\hat{g}\cos (N\hat{\vphi})$, respectively, with $g, \hat{g}\gg 1$. 

To obtain the twisted smooth boundary, we couple the boundary to the double semion model in equation \eqref{eq:ds} and focus on $N=2$. 
We couple the 3d and 2d theories such that the total boundary Lagrangian is, 
\be
\begin{split}
\L_{\text{tot}}=\frac{2}{2\pi}\epsilon_{ij} \big[(\p_i\vphi_j) \p_0\hat{\vphi}& + (A_{1i} \p_0 A_{2j}+\frac{1}{2}A_{i2}d A_{2j}) \big] \\
& + \lambda_{i} (A_{2i} - \p_i \hat{\vphi}). 
\end{split}
\label{eq:couple_ds}
\ee
where we have used chosen the temporal gauge for $A_1$ and $A_2$, and 
$\lambda$ is a Lagrange multiplier that physically identifies the charge excitation $e$ on the boundary of the toric code and the nonchiral anyon $s\bar{s}$ in double semion. We have also chosen the 2d surface to live on $T^2$. 
Imposing the equation of motion for $\lambda_i$, we obtain $A_{2i}=\p_i\hat{\vphi}$ and the Lagrangian reduces to
\be
\L_{\text{eff}}=\frac{2}{2\pi} (\p_0\hat{\vphi}) \epsilon_{ij}\p_i\left(\vphi_j +A_{1j}\right)
\label{eq:effective_ds}
\ee
The gapping term is therefore $-g \sum_j \cos (2 \vphi_j + 2 A_{2j})$. The vertex operators get modified 
\be
\tilde{V}_j =e^{i\oint dx_j(\vphi_j+A_{1j})},
\ee
corresponding to the dressing of the endpoint of $m$-string by a semion, with the semionic statistics visible from 
\be
\tilde{V}_i(x_j')\tilde{V}_j(x_i') = e^{i\pi \epsilon_{ij}}\tilde{V}_j(x_i')\tilde{V}_i(x_j). 
\ee
The above equation directly follows from integrating out $A_2$ in the double semion Lagrangian \eqref{eq:ds}, leading to
\be
\tilde{\L}\rightarrow -\frac{2}{4\pi} A_1 dA_1. 
\ee

Turning back to the more general case, we can add to \eqref{eq:boundary_phi} the boundary Lagrangian of the 2d $\Z_{\tilde{N}}$ toric code 
\be
\tilde{\L}=\frac{\tilde{N}\sigma_x}{4\pi} A_I dA_J.
\ee 
We can again take the temporal gauge and add a term that couples the two theories, $\lambda (A_{2i}-\p_i\hat{\vphi}).$
Upon plugging in the equation of motion for $\lambda,$ we get
\be
\L_{\text{eff}}= \frac{N}{2\pi}\left[(\p_0\hat{\vphi}) \epsilon_{ij}\p_i \left(\vphi_j+n A_{1j}\right)\right]. 
\label{eq:N_tilde}
\ee
Here $n\equiv \tilde{N}/N$. 
When $n=1$, we can recover the smooth boundary by adding $-g \sum_j \cos(N\vphi_j +N A_{1j})$ with a large $g$. Notice that in this case, $A_1$ trivially commutes with itself. Similarly, the exchange smooth boundary can be obtained by exchanging $A_1\leftrightarrow A_2$ in the coupling as well as the gapping terms. 

Generally $n\neq 1$. For convenience, we label the anyons in the 2d $\Z_{\tilde{N}}$ topological order as $\tilde{e}^a \tilde{m}^b$ with $a, b\in \Z$ mod $\tilde{N}$. The effective Lagrangian \eqref{eq:N_tilde} then shows that the charge excitation $e$ from the 3d $\Z_N$ toric code on the boundary has mutual semionic statistics with the combination of: boundary flux excitation $m$ from the 3d $\Z_N$ toric code and the anyon $e^0 m^{n}$. This clearly does not make sense when $n$ is not an integer, therefore the only sensible choice is to add $-\hat{g}\cos (N\hat{\vphi})$ and arrive at a rough-type boundary. Therefore, only when $n\in \Z$, the condensation of pairs of charges in the 3d and 2d toric codes allows the $m$-string to end on the boundary, and the corresponding boundary belongs to the smooth boundary class. 

Alternatively some other coupling terms can be designed such that $e$ from the 3d toric code condenses with a generic $e^a m^b$ in the 2d toric code, as long as $e^am^b$ is a boson. The simplest nontrivial choice is $e^2m^2$ when $\tilde{N}=4,$ and the $m$-string can have endpoints $e^0m^1$ or $e^1m^0$. There are also deconfined $2d$ excitations $e^1m^1$ and $e^2$. The condensation of $e^2m^2$ in $\Z_4$ 2d toric code was previously discussed in ref. \cite{Ellison_2022}. 

\subsection{Anomaly inflow}
\label{subsec:inflow}

In this section, we use the boundary theory \eqref{eq:N_tilde} with $n\equiv (\tilde{N}/N) \in \Z$ and anomaly inflow  \cite{CALLAN1985427} to construct the corresponding bulk (3+1)d theory such that the bulk and boundary anomalies cancel (see for example \cite{anomaly,Me} for procedures in similar contexts).

We start with the $n=0$ case when no (2+1)d phase is added and couple the symmetries of the boundary theory \eqref{eq:phi_sym} to the following $1$-form and $2$-form $U(1)$ background fields
\be
\hat{B}_{\mu}\sim \hat{B}_{\mu}+\p_{\mu}\hat{\alpha},\ B_{\mu\nu}\rightarrow B_{\mu\nu}+(\p_{\mu}\alpha_{\nu}-\p_{\nu}\alpha_{\mu}). 
\label{eq:added_2form}
\ee
The boundary Lagrangian after coupling is
\be
\L'=\frac{N}{2\pi}(\p_0\hat{\vphi}) \epsilon_{ij}\p_i\vphi_j - N(\hat{B}_0 J_0+B_{0j}\hat{J}_{0j}).
\ee
One can also add local counterterms which do not affect the anomaly but will make the expression look nicer,
\be
\L_{\text{ct}}=\frac{N}{4\pi}(\hat{B}_0 B_{xy}+\epsilon_{ij}\hat{B}_i B_{0j}). 
\ee
Under a gauge transformation of the added $B$ and $\hat{B}$ fields, as well as the corresponding transformations for $\vphi_j$ and $\hat{\vphi}$, the Lagrangian is not invariant and changes as
\be
\delta(\L'+\L_{\text{ct}})=-\frac{N}{4\pi}\epsilon_{z\mu\nu\rho}\left(\hat{\alpha} \p_\mu B_{\nu \rho}+ \alpha_{\mu}\p_{\nu} \hat{B}_\rho \right).
\label{eq:anomaly}
\ee
This cannot be removed by any local counterterm and the boundary theory thus has mixed 't Hooft anomaly between the two symmetries. 
 
This anomaly can however be canceled by the following $(3+1)$d theory constructed from the $B$ and $\hat{B}$ fields:
\be
\L_{3+1}=\frac{N}{2\pi} Bd\hat{B}.
\ee
Under gauge transformations, $\L_{3+1}$ changes as a boundary term,
\be
\delta\L_{3+1}=\frac{N}{4\pi} \epsilon_{z\nu\rho\sigma} \p_z  (\hat{\alpha}\p_{\mu}B_{\nu\rho}+\alpha_{\mu}\p_{\nu} \hat{B}_{\rho}),
\ee
which exactly cancels the anomaly in \eqref{eq:anomaly}. We would like to comment that while $\L_{3+1}$ has the same form as a (3+1)d toric code theory, the $B$ and $\hat{B}$ fields are not dynamical gauge fields but background ones. $\L_{3+1}$ is a symmetry protected topological phase, which can be obtained from the true toric code theory by coupling toric code to corresponding background fields, and integrating out the dynamical fields. 

Next we move on to the case when $n\neq 0$ is a finite integer. 
There are additional gauge symmetries due to the presence of $A_1$: one is the usual gauge symmetry of $A_1\sim A_1+d\gamma$, the other is $
\vphi_j\sim \vphi_j+\lambda_j(t,\bx),$ $ A_{1j}\sim A_{1j}-\lambda_j(t,\bx).$
We also have the additional $\Z_{\tilde{N}}$ one-form symmetry due to $A_1$, generated by
\be
\tilde{W}(\mathcal{C})=\exp(i\oint_{\mathcal{C}} A_1).
\ee
One thus needs to introduce additional $\Z_{\tilde{N}}$ 2-form background field. We do this by first introducing $U(1)$ 2-form field $C$ and then using an additional dynamical compact scalar field $\phi$ to constrain $C$ to be $\Z_{\tilde{N}}$ 2-form field. We choose the gauge transformations of $C$ to be
\be
A\rightarrow A+\beta,\quad C\rightarrow C+d\beta, 
\ee
where the one-form gauge parameter $\beta$ has its own gauge symmetry $\beta\sim \beta+d\xi$. The dynamical $U(1)$ gauge symmetry of $A_1$ acts on $\phi$ as $\phi\sim \phi+\gamma.$

The boundary Lagrangian after coupling is
\be
\begin{split}
\L'=\frac{N}{2\pi}& \epsilon_{ij}\big[(\p_0\hat{\vphi}-B_0) \p_i(\vphi_j + n A_{1j})\\
& -(B_{0j}+nC_{0j})\p_i\hat{\vphi}\big]-\frac{\tilde{N}}{2\pi}\phi dC.
\end{split}
\ee
One again adds local counterterms to make the expression look nicer,
\be
\L_{\text{ct}}=\frac{N}{4\pi}[\hat{B}_0 (B_{xy}+nC_{xy})+\epsilon_{ij}\hat{B}_i (B_{0j}+nC_{0j})]. 
\ee
Under gauge transformations, the Lagrangian $\L'+\L_{\text{ct}}$ is not invariant and changes as 
\be
\begin{split}
\delta(\L'+\L_{\text{ct}})=  -& \frac{N}{4\pi} \epsilon_{z\mu\nu\rho}\big[\hat{\alpha} \p_\mu (B_{\nu \rho}+C_{\nu \rho})\\
& + (\alpha_{\mu}+\beta_{\mu})\p_{\nu} \hat{B}_\rho +n\gamma\p_{\mu}C_{\nu\rho}\big].
\end{split}
\label{eq:anomaly_twisted}
\ee
This cannot be removed by any local counterterm, signaling the 't Hooft anomaly. 
 
This anomaly can however be easily canceled by the following $(3+1)$d bulk: 
\be
\L_{3+1}=\frac{N}{2\pi} [(B+nC) d\hat{B}+n\hat{C}dC],
\ee
where $\hat{C}\sim \hat{C}+d\gamma$ is a dynamical one-form gauge field that constraints $C$ to $\Z_{\tilde{N}}$.

\section{3d bTC coupled to exciton bose liquid} 
\label{sec:EBL}

In this section we examine the case when the added 2d phase of matter is not a topological order, but a tensor gauge theory. We will focus on the $\Z_N$ plaquette model \cite{10.21468/SciPostPhys.10.2.027}, which can be obtained by coupling the $U(1)$ exciton Bose liquid \cite{EBL} to a scalar field $N$ that Higgses it to $\Z_N$. 
For convenience we will focus on the $N=2$ case, but the generalization to arbitrary integer $N$ is straightforward and amounts to using generalized Pauli operators and keeping track of the hermitian conjugates.  We will discuss this 3d $\Z_2$ toric code plus 2d $\Z_2$ plaquette model using both lattice and field theories. 

\subsection{Lattice description}
We first briefly review the $\Z_N$ plaquette model: the qubits live on the vertices of a square lattice, and the Hamiltonian is 
\be
H_{\text{plaq}}=-K\sum_{x,y}  \tilde{Z}_{x,y} \tilde{Z}_{x+1,y} \tilde{Z}_{x,y+1} \tilde{Z}_{x+1,y+1}. 
\label{eq:EBL_ham}
\ee
The conserved charges are 
\be
W_x(x)=\prod_y \tilde{X}_{x,y},\quad W_y(y)=\prod_x \tilde{X}_{x,y}.
\ee

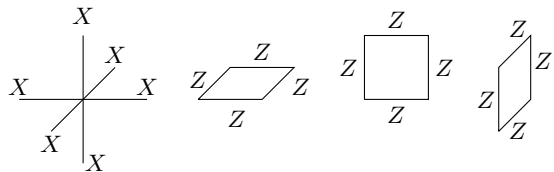
\begin{figure}[htbp]
    \centering
    \begin{tikzpicture}[scale=0.85]
    \draw (0,0)--(2,0);
    \draw (1,0)--(1.5,0.5);
    \draw (1,0)--(0.5,-0.5);
    \draw (1,0)--(1,-1);
    \node at (0,0.2) {$X$};
    \node at (2,0.2) {$X$};
    \node at (0.5,-0.7) {$X$};
    \node at (1.5,0.7) {$X$};
    \node at (1,1.3) {$X$};
    \node at (1.2,-1) {$X$}; 
    \draw (1,0)--(1,1);
    \draw (3-0.2,0)--(4-0.2,0)--(4.3,0.5)--(3.3,0.5)--(2.8,0);
    \node at (3.4,-0.3) {$Z$};
    \node at (4.4,0.2) {$Z$};
    \node at (2.8,0.3) {$Z$};
    \node at (3.8,0.75) {$Z$};
\draw (5+0.4,0)--(6+0.4,0)--(6+0.4,1)--(5+0.4,1)--(5+0.4,0);
\node at (5.15,0.5) {$Z$};
\node at (6.65,0.5) {$Z$};
\node at (6-0.1,1.25) {$Z$};
\node at (6-0.1,-0.25) {$Z$};
\draw (7+0.5,0-0.5)--(7.5+0.5,0)--(7.5+0.5,1)--(7+0.5,0.5)--(7+0.5,-0.5);
\node at (7.3,0) {$Z$};
\node at (8.2,0.6) {$Z$};
\node at (7.8,-0.5) {$Z$};
\node at (7.8,1.1) {$Z$};
\end{tikzpicture}
\caption{Hamiltonian terms of the 3d $\Z_N$ toric code on the cubic lattice, without boundaries. }
    \label{fig:cubic_TC}
\end{figure}

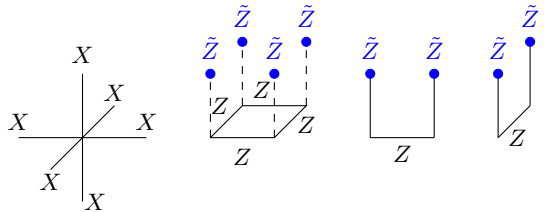
\begin{figure}[htbp]
    \centering
    \begin{tikzpicture}[scale=0.85]
    \draw (0,0)--(2,0);
    \draw (1,0)--(1.5,0.5);
    \draw (1,0)--(0.5,-0.5);
    \draw (1,0)--(1,-1);
    \node at (0,0.25) {$X$};
    \node at (2,0.25) {$X$};
    \node at (0.5,-0.7) {$X$};
    \node at (1.5,0.7) {$X$};
    \node at (1,1.3) {$X$};
    \node at (1.2,-1) {$X$}; 
    \draw(1,0)--(1,1);
    \draw (3,0)--(4,0)--(4.5,0.5)--(3.5,0.5)--(3,0);
    \draw[dashed] (3,0)--(3,1);
    \draw[dashed] (4,0)--(4,1);
    \draw[dashed] (4.5,0.5)--(4.5,1.5);
    \draw[dashed] (3.5,0.5)--(3.5,1.5);
    \filldraw[blue] (3,1) circle (2pt);
    \filldraw[blue] (4,1) circle (2pt);
    \filldraw[blue] (4.5,1.5) circle (2pt);
    \filldraw[blue] (3.5,1.5) circle (2pt);
    \node at (3.5,-0.3) {$Z$};
    \node at (4.5,0.2) {$Z$};
    \node at (3.15,0.5) {$Z$};
    \node at (3.8,0.75) {$Z$};
    \node at (3.5,1.9) {\textcolor{blue}{$\tilde{Z}$}};
    \node at (4.5,1.9) {\textcolor{blue}{$\tilde{Z}$}};
    \node at (3,1.4) {\textcolor{blue}{$\tilde{Z}$}};
    \node at (4,1.4) {\textcolor{blue}{$\tilde{Z}$}};
\draw 
(6+0.5,1-1)--(5+0.5,1-1);
\draw (5+0.5,1-1)--(5+0.5,2-1);
\draw (6+0.5,1-1)--(6+0.5,2-1);
\filldraw[blue] (5+0.5,2-1) circle (2pt);
\filldraw[blue] (6+0.5,2-1) circle (2pt);
\node at (6,-0.25) {$Z$};
\node at (5.5,1.4) {\textcolor{blue}{$\tilde{Z}$}};
\node at (6.5,1.4) {\textcolor{blue}{$\tilde{Z}$}};
\draw (7.5+0.5,1.5-1)--(7+0.5,1-1);
\draw (7+0.5,1-1)--(7+0.5,2-1);
\draw (7.5+0.5,1.5-1)--(7.5+0.5,2.5-1);
\filldraw[blue] (7+0.5,2-1) circle (2pt);
\filldraw[blue] (7.5+0.5,2.5-1) circle (2pt);
\node at (7.8,0) {$Z$};
\node at (7.5,1.4) {\textcolor{blue}{$\tilde{Z}$}};
\node at (8,1.9) {\textcolor{blue}{$\tilde{Z}$}};
\end{tikzpicture}
\caption{Modified Hamiltonian on the surface. The dashed lines are only aids for the eye.}
\label{fig:EBL_bTC}
\end{figure}

We couple it to the 3d toric code. The 3d toric code on the cubic lattice has the Hamiltonian 
\be
H_{TC}=-\sum_v \prod_{i\in v} X_i-\sum_p \prod_{i\in p} Z_i,
\ee
with $v$ labeling vertices and $p$ labeling plaquettes, see  fig. \ref{fig:cubic_TC}. We have chosen a basis  different from that in section \ref{sec:WW}.
Near the (top) surface, the terms get modified as shown in figure \ref{fig:EBL_bTC}. The vertex terms remain intact, while additional $\tilde{Z}$'s are attached to the plaquette terms near the surface.  

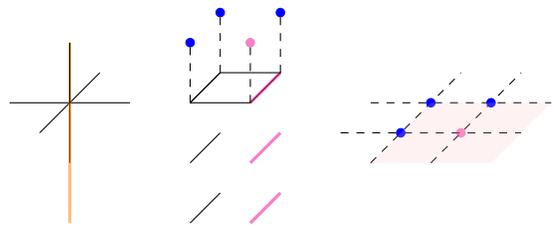
\begin{figure}[htbp]
    \centering
    \begin{tikzpicture}[scale=0.8]
    \draw (0,0)--(2,0);
    \draw (1,0)--(1.5,0.5);
    \draw (1,0)--(0.5,-0.5);
    \draw (1,0)--(1,-1);
    \draw[opacity=0.5,orange,very thick] (1,-2)--(1,1);
    \draw (1,0)--(1,1);
    \draw (3,0)--(4,0)--(4.5,0.5)--(3.5,0.5)--(3,0);
    \draw[magenta, very thick, opacity=0.5] (4,0)--(4.5,0.5);
    \draw[magenta, very thick, opacity=0.5] (4,0-1)--(4.5,0.5-1);
    \draw[magenta, very thick, opacity=0.5] (4,0-2)--(4.5,0.5-2);
    \draw (4-1,0)--(4.5-1,0.5);
    \draw (4-1,0-1)--(4.5-1,0.5-1);
    \draw(4-1,0-2)--(4.5-1,0.5-2);
    \draw[dashed] (3,0)--(3,1);
    \draw[dashed] (4,0)--(4,1);
    \draw[dashed] (4.5,0.5)--(4.5,1.5);
    \draw[dashed] (3.5,0.5)--(3.5,1.5);
    \filldraw[blue] (3,1) circle (2pt);
    \filldraw[magenta,opacity=0.5] (4,1) circle (2pt);
    \filldraw[blue] (4.5,1.5) circle (2pt);
    \filldraw[blue] (3.5,1.5) circle (2pt);
    \draw[dashed] (6,0)--(9,0);
    \draw[dashed] (5.5,-0.5)--(8.5,-0.5);
    \draw[dashed] (6,-1)--(7.5,0.5);
    \draw[dashed] (7,-1)--(8.5,0.5);
    \filldraw[blue] (7,0) circle (2pt);
    \filldraw[blue] (8,0) circle (2pt);
    \filldraw[blue] (7-0.5,0-0.5) circle (2pt);
    \filldraw[magenta, opacity=0.5] (8-0.5,0-0.5) circle (2pt);
    \fill[fill=pink, opacity=0.2] (7,0)--(8,0)--(7,-1)--(6,-1)--(7,0);
    \fill[fill=pink, opacity=0.2] (8,0)--(9,0)--(8,-1)--(7,-1)--(8,0);
    \end{tikzpicture}
    \caption{Left: acting $Z$ on the orange links moves a charge $e$ from bulk to the vacuum. Middle: acting $X$ on the pink y-links and pink vertex creates a $m$ string whose endpoint lives in the $\Z_N$ plaquette layer. Right: More details on the top $\Z_N$ plaquette layer.  Endpoint of this m-string corresponds to a quadruple plaquette excitations.}
    \label{fig:EBL_mobility}
\end{figure}

Starting from the ground state, consider acting $Z$ on an orange path extending from the bulk to the surface (leftmost panel of figure \ref{fig:EBL_mobility}), such that a bulk electric charge $e$ moves toward the surface and vanishes into the vacuum. 
One can also consider a magnetic string that extends from the bulk towards the surface. This string can end on the top surface by acting $\tilde{X}$ on the pink vertex in figure \ref{fig:EBL_mobility}. The endpoint corresponds to a quadrupole of plaquette excitations. A dipole of plaquette excitations can move in one spatial dimension but only at the cost of creating additional excitations in the bulk, as the $m$-string costs an energy proportional to its length.

\subsection{Field theory} 
\label{subsec:EBL_field}
The field theory for the (2+1)d $\Z_N$ plaquette model is \cite{10.21468/SciPostPhys.10.2.027}, 
\be
\L_{\text{plaq}}=\frac{N}{2\pi} \phi^{xy}(\p_0 A_{xy}-\p_x\p_y A_0),
\label{eq:Lagrangian_EBL}
\ee
with the gauge transformations $A_0\sim A_0+\p_0\beta,$ $A_{xy}\sim A_{xy}+\p_x\p_y\beta$. 
The generator of the $\Z_N$ electric global symmetry is $\exp [i\phi^{xy}(x,y)]$, while the generators of the $\Z_N$ dipole global symmetry are
\be
\begin{split}
& W_x(x_1,x_2)=\exp \left[i\int_{x_1}^{x_2} dx\oint dy A_{xy}\right],\\ 
& W_y(y_1,y_2)=\exp \left[i\int_{y_1}^{y_2} dy\oint dx A_{xy}\right].
\end{split}
\label{eq:tensor}
\ee
They satisfy the $L_x+L_y-1$ copies of $\Z_N$ Heisenberg algebra, 
\be
\begin{split}
& e^{i\phi^{xy}(x,y)}W_x(x_1,x_2)=e^{2\pi i/N}W_x(x_1,x_2)e^{i\phi^{xy}(x,y)},\\
& e^{i\phi^{xy}(x,y)}W_y(y_1,y_2)=e^{2\pi i/N}W_y(y_1,y_2)e^{i\phi^{xy}(x,y)},\end{split}
\ee
when $x_1<x<x_2$ and $y_1<y<y_2$. Notice that while the $\Z_N$ dipole symmetry is present on the lattice, the $\Z_N$ electric symmetry is easily broken by 
a small $- h\sum_{x,y} \tilde{X}_{x,y}$ term that can be added to the \eqref{eq:EBL_ham} without inducing a phase transition. 
From the Lagrangian \eqref{eq:Lagrangian_EBL}, one can see the density operators 
\be
\hat{\mathcal{\rho}}=\frac{1}{2\pi}\phi^{xy},\quad {\rho}=\frac{1}{2\pi} A_{xy}. 
\ee
The corresponding quantized charges are,
\be
\begin{split}
\int_{y_1}^{y_2}dy\oint dx\  \rho =m_x(y_1,y_2),\\
\int_{x_1}^{x_2}dx \oint dy\ \rho=m_y(x_1,x_2),\\
\hat{\rho}=m(x,y),\quad m,\ m_i\in \mathbb{Z}. 
\end{split}
\ee
We couple the gapless boundary theory of the 3d toric code \eqref{eq:boundary_phi} to the $\Z_N$ plaquette model by adding
\be
\L=\frac{N}{2\pi} (\p_0\hat{\vphi}) \epsilon_{ij}\p_i \vphi_j+{\lambda} (\hat{\vphi}-\phi^{xy}).
\ee
Imposing the equations of motion for $\lambda$ 
and choosing the temporal gauge for the plaquette model, the total boundary theory reduces to
\be
\L_{\text{eff}}= \frac{N}{2\pi}[ (\p_0\hat{\vphi}) (\epsilon_{ij}\p_i\vphi_j - A_{xy})],
\label{eq:EBL_total}
\ee
and the gapping term can be chosen $-g\cos(N\epsilon_{ij}\p_i\vphi_j-NA_{xy}),$ corresponding to the fact that the endpoint of the $m$-string is now dressed by the plaquette violation operators in the $\Z_N$ plaquette model. This boundary theory belongs to the smooth class of gapped boundaries of toric code.

Now we study the anomaly inflow for this boundary theory \eqref{eq:EBL_total}. With the presence of $A_{xy}$, we take into account of the higher rank tensor symmetries \eqref{eq:tensor} and couple them to the following $U(1)$ tensor gauge field $C_0^{xy}\sim C_0^{xy} +\p_0\beta_{xy}$.
The coupled Lagrangian is 
\be
\begin{split}
\L'= & \frac{N}{2\pi}\big[ (\p_0\hat{\vphi}-\hat{B}_0)(\epsilon_{ij}\p_i\vphi_j-A_{xy})-B_{0j}(\epsilon_{ij}\p_i\hat{\vphi})\\
& -\hat{\vphi}C_0^{xy}\big]+\frac{1}{2\pi}\big[ \chi (\dot{\psi}-NC_0^{xy})\big],
\end{split}
\ee
where $\hat{B}_0$ and $B_{0j}$ were defined in \eqref{eq:added_2form},  $\chi$ is a real Lagrange multiplier and $\psi$ is circle-valued field that changes under gauge transformation as $\psi\sim \psi+N\beta_{xy}.$ The $\chi$ term Higgses the holonomy of $C_0^{xy}$ down to $\Z_N$. We also add the local counterterms exactly as in the case without $A_{xy}$, 
\be
\begin{split}
\L_{\text{ct}}=\frac{N}{4\pi}(\hat{B}_0B_{xy}+\epsilon_{ij}\hat{B}_iB_{0j}).
\end{split}
\ee
The total change of the Lagrangian under gauge transformations of $B$, $\hat{B}$ and $C$ are
\be
\begin{split}
\delta (\L'+\L_{\text{ct}})=& -\frac{N}{4\pi}\epsilon_{z\mu\nu\rho}\left(\hat{\alpha} \p_\mu B_{\nu \rho}+ \alpha_{\mu}\p_{\nu} \hat{B}_\rho \right)\\
& \ \ +\frac{N}{2\pi}[\beta_{xy}(\hat{B}_0+\p_0\hat{\alpha})-\hat{\alpha}C_0^{xy}]. 
\end{split}
\ee
This anomaly can be canceled by the following $(3+1)$d theory with a $z=0$ boundary, 
\be
\begin{split}
\L_{3+1}=\frac{N}{2\pi}&\big[Bd\hat{B}+ \Psi_{xy} (\p_0\hat{B}_z-\p_z\hat{B}_0)+\hat{B}_z C_0^{xy}\\
&  -\hat{B}_0 C_z^{xy} -\hat{\Psi}(\p_0 C_z^{xy}-\p_z C_0^{xy})\big],
\end{split}
\ee
where $\hat{\Psi}\sim \hat{\Psi}+\hat{\alpha}$ and $\Psi_{xy}\sim \Psi_{xy}+\beta_{xy}$ are two dynamical fields that Higgs the $U(1)$ fields down to $\Z_N$. $C_z^{xy}\sim C_z^{xy}+\p_z \beta_{xy}$ and $\hat{B}_z\sim \hat{B}_z +\p_z\hat{\alpha}$ are additional components of the previously defined gauge fields in the added spatial dimension.

\section{Generalizations and Discussions}
\label{sec:discussion}

In this section, we comment on the generalization to other 3d topological orders and future directions. To this end, it is the easiest to use the Walker-Wang plus string-net formalism. Consider the input of the Walker-Wang model to be a unitary fusion category, in particular, the representation category of a finite group $G$. For convenience, we focus on the untwisted case (the trivial element in $H^3[G,U(1)]$). The bulk excitations thus include the string-like flux excitations corresponding to the conjugacy classes of $G$, with endpoints being pointlike charges  labeled by the irreducible representations of the centralizer of the conjugacy class. When the conjugacy class is trivial, the excitation just reduces to pointlike particles corresponding to the irreducible representations of $G$. 

When a boundary is present, we can again use the similar picture as in fig. \ref{fig:idea}, but with $e$ substituted by a bosonic charge $q$, which pairs up and condenses together with a boson $b$ in the added 2d TO. In general there can be multiple pairs of $\{(q_i, b_i)\}$ which together form a closed set $S$ of commuting composites that are allowed to condense in a compatible way. By closed, we mean that if $(q_1, b_1)\in S$ and $(q_2, b_2)\in S$, then $(q_1\times q_2, b_1\times b_2)\in S$. In other words, the fusion between vacuum and vacuum should not be nontrivial. We do not require the set to be maximal, i.e. the theory can contain additional composites $\{(\tilde{q}_j, \tilde{b}_j)\}$ that commute with the elements $S$ but $(\tilde{q}_j, \tilde{b}_j)\notin S$. Consequently, while the composites $q_i\otimes b_i$ become the new vacuum on the boundary,  $\tilde{q}_j\otimes \tilde{b}_j$ is still a well-defined excitation. 

As for the various types of string excitations $p_{\alpha}$'s in the bulk, they again can only end at the boundary if there exists at least one $f$ in the 2d TO such that its monodromy with each $b_i$ cancels the corresponding monodromy between $p_{\alpha}$ and $q_i$. Then different classes of gapped boundaries are organized by which types of string $p_{\alpha}$'s can end on the boundary. In the case where the bulk theory is characterized by Rep$_G$ the representation category of a finite group, the different classes of gapped boundaries correspond to the different normal subgroups of $G$, ignoring all possible complications arising from 3-cochains in $C^3[K,U(1)]$, and the possible couplings to many complicated 2d topological orders. 

Notice that our example of 3d fermionic toric code described in \ref{subsec:layer_fTC} is already beyond the discussion in the last paragraph, where instead of condensing a pair of bosons, a pair of fermions is condensed. In general as long as the composite is a boson, the condensation is allowed. 

We would like to comment on the concept of ``classes'' of gapped boundaries for 3d TOs. We arrive at different types of boundaries of the 3d TO by coupling the bulk theory to an additional 2d exotic phase of matter (including TO and fractonic theories). Each elementary type of gapped boundary is characterized by the possible surface excitations, their statistics and mobility constraints. By elementary, it means that we mod out any 2d surface theory that is decoupled from the bulk. 
Then we distinguish different classes of gapped boundaries by whether different types of string operators can end at the boundary, regardless of the potentially different endpoints. As also reviewed in section \ref{subsec:previous}, another different, mathematically natural way to understand the gapped boundaries is to use Lagrangian algebras of the modular 2-categories, based on which there should be three classes for the $\Z_2$ toric code case \cite{Janet}. It is unfortunate that in our method, it is not natural to single out the twisted smooth boundary condition as another third class.

It would be interesting to examine the case where the input category of the 3d Walker-Wang model is that of a modular tensor category, such that no nontrivial excitations exist in the bulk, but there can be nontrivial surface excitations. One can then consider condensing the surface excitations together with some anyons in an added 2d topological order to find exotic boundaries. Since the idea is the same as illustrated in the previous sections, we will not go into further details, but would like to point out that previously there have been discussions on the surface topological orders of symmetry protected topological phases, see for example \cite{PhysRevX.5.041013,1804.08628,PhysRevX.3.041016,PhysRevLett.112.231602,PhysRevB.89.235103,PhysRevX.8.031048,anomaly}. 

The criterion for different classes of 2d gapped boundaries of 3d TOs in this work also motivates the following generalization to higher dimensions: different classes of the (n-1)d gapped boundaries of nd topological orders may be characterized by the different (n-2)d excitations that can end on the boundary. More future work will be needed to validate or invalidate this conjecture. 

\acknowledgments This project was initially motivated by a discussion with Shu-Heng Shao on a related fracton work \cite{PhysRevB.106.195102}. We are grateful to Ashvin Vishwanath for many insightful comments and questions, and to Ho-Tat Lam for important suggestions on the draft. We thank Yuting Hu,  Yidun Wan and Juven Wang for explaining their previous works \cite{Full,Yidun,Juven}; Meng Cheng, Po-Shen Hsin, Chao-Ming Jian, Nathanan Tantivasadakarn, Ryan Thorngren and Carolyn Zhang for helpful conversations; and Cenke Xu for informing us of their related work \cite{Coordinate} on deconfined quantum critical points which will appear in the same arXiv listing. This work is supported by the Simons Collaborations on Ultra-Quantum Matter, grant 651440 (S.S. and A.V.) from the Simons Foundation. After posting the first version of this manuscript, a related work was brought to our attention \cite{PoShen} which will appear on arXiv soon. 


\bibliography{ref}

\appendix
\begin{widetext}
    
\section{Contraction of charges in the string-net model}
\label{app:contraction}

In this appendix, we review the procedure of charge contraction in order to construct dyon loops out of shorter dyon strings. The materials are summaries and applications of the results in ref. \cite{Full}. 

We first introduce the rules for the charge tails to move:
\be
 \vcenter{\hbox{\includegraphics[scale = 0.17]{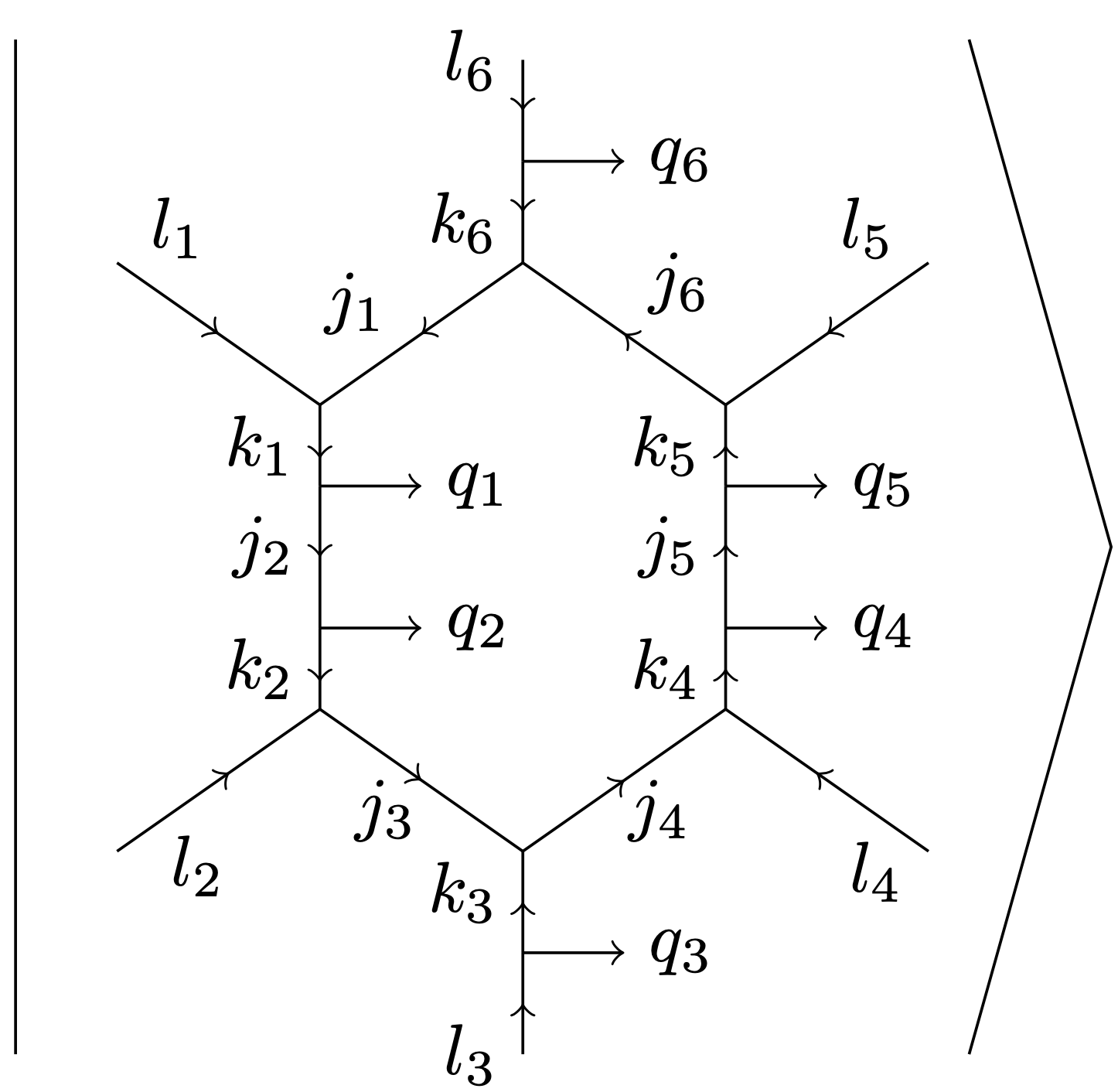}}}
\rightarrow \sqrt{\frac{d_{k_1}}{d_{k_1'}}} G_{j_1l_1k_1'}^{j_2^*q_1^*k_1}\vcenter{\hbox{\includegraphics[scale = 0.24]{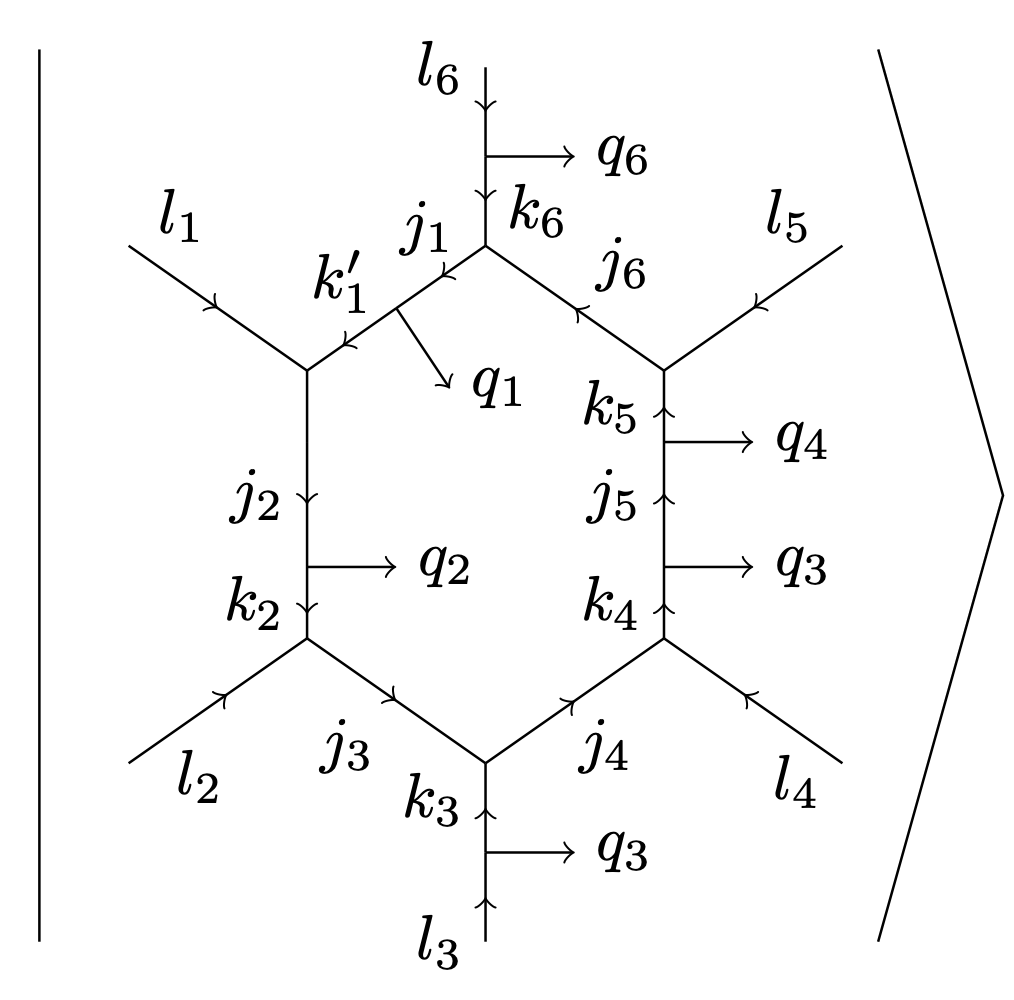}}}
\label{eq:move_charge}
\ee
The tails can move around freely while remaining on the same side of the plaquette, unless bumping into another charge tail on the same side. The move in \eqref{eq:move_charge} resembles the elementary F- or the recoupling-move of the graph in the original string-net model. 

Equipped with the adequate tool, we can now create the short loop of $K$ dyon as described in section \ref{sec:WW}. We start by creating three short strings near the three edges surrounding a vertex, 
\be
\begin{split}
& W_{e_3}^{K;q_1^*q_3}W_{e_2}^{K;q_2q_3^*} W_{e_1}^{K;q_1q_2^*}\vcenter{\hbox{\includegraphics[scale = 0.25]{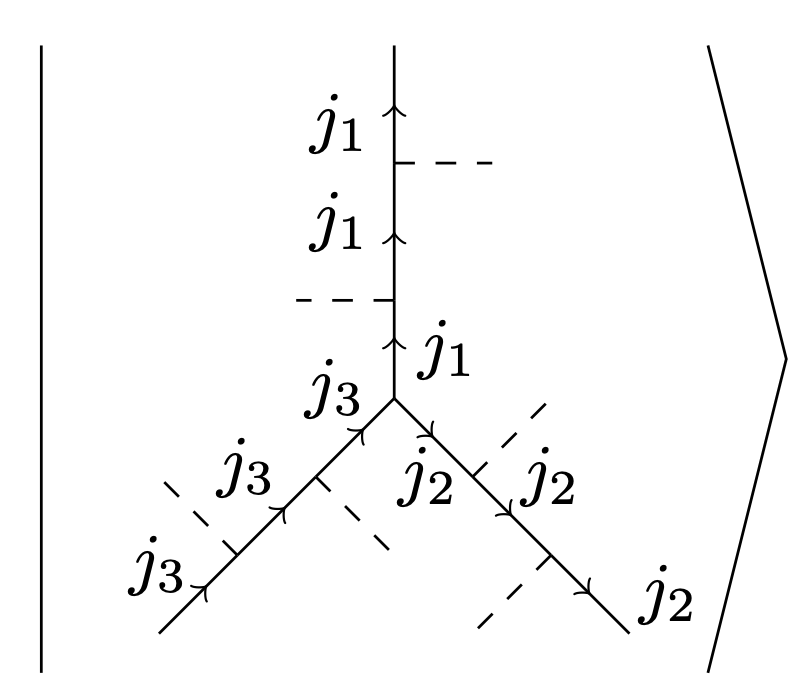}}}\\
= & \sum_{q_1,q_2,q_3}\sum_{j_1',j_2',j_3'}\sqrt{\frac{d_{j_1'} d_{j_2'} d_{j_3'}}{d_{j_1} d_{j_2} d_{j_3}}} \overline{z_{q_1^*j_3'q_3^*j_3}^K} \overline{z_{q_2^*j_2'q_3^*j_2}^K}\overline{z_{q_1j_1'q_2j_1}^K}
\vcenter{\hbox{\includegraphics[scale = 0.25]{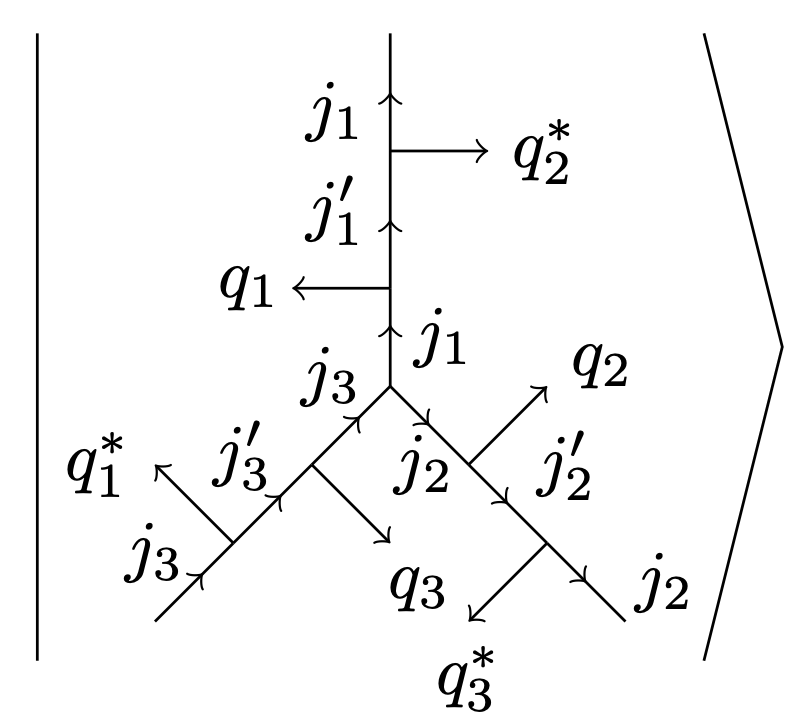}}}
\end{split}
\label{eq:create_loop}
\ee
Next we use the move introduced in \eqref{eq:move_charge} to place $q_i$ and $q_i^*$ on the same link, and annihilate each pair by
\be
\mathcal{P}:\quad  \vcenter{\hbox{\includegraphics[scale = 0.25]{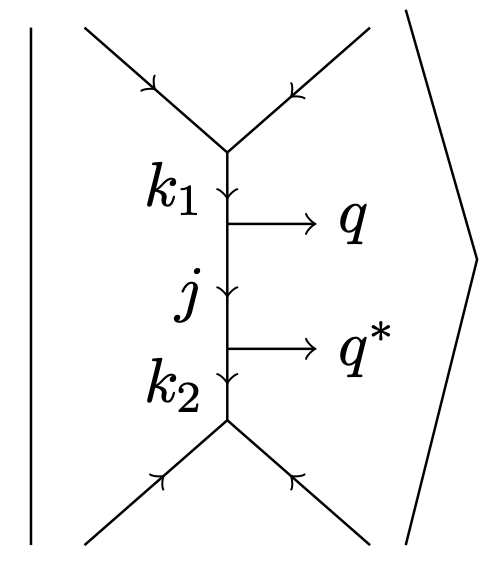}}}
\rightarrow \delta_{k_1 k_2} \delta_{q k_1 j_2^*} (d_j/d_{k_1})^{1/2}\vcenter{\hbox{\includegraphics[scale = 0.25]{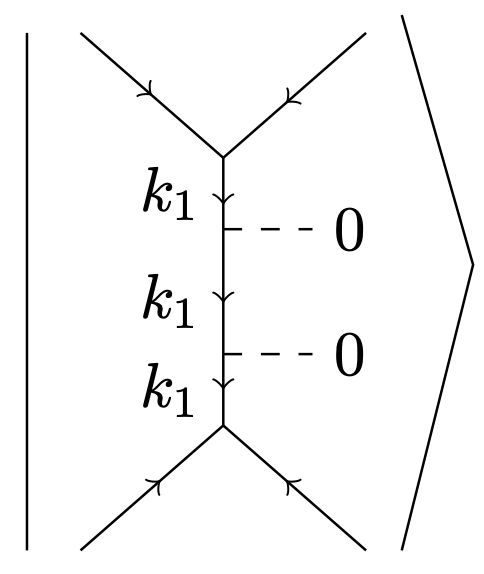}}}.
\label{eq:project_charge}
\ee
When the initial state is one without any charges, as in \eqref{eq:create_loop}, after some simplifications using the symmetry properties of $z$ and $G$, this whole procedure simply gives
\be
O_v^K \vcenter{\hbox{\includegraphics[scale = 0.25]{fig_create_loop1.png}}} = \sum_{q_1,q_3} \sum_{j_3'} \frac{d_{j_3'}}{d_{j_3}}z^{K}_{q_1j_3q_3j_3'} \overline{z^K_{q_1^*j_3'q_3^*j_3}} \vcenter{\hbox{\includegraphics[scale = 0.25]{fig_create_loop1.png}}},
\ee
corresponding to shrinking the short loop $K$ such that it crosses the edge $e_3$ twice but does not cross the other two edges $e_1$ or $e_2$. 
However, when the initial state is not the ground state and has nontrivial charge excitations near the vertex, the newly generated charge tails can no longer freely move around.

\end{widetext}

\end{document}